\documentclass{JHEP3}
\usepackage{amsmath}
\usepackage{epsfig}

\newcommand{\dint} {{\mbox{\reflectbox{${\displaystyle \int}$}}}}

\newcommand{\Ztwo}{{\mathbb{Z}_2}}

\newcommand{\One} {{\bf 1}} 

\newcommand{\tQ}{{(Q-\eta_0)}}

\newcommand{\cA} {{\cal A}}
\newcommand{\cL} {{\cal L}}

\newcommand{\cO} {{\cal O}}
\newcommand{\cQ} {{\cal Q}}
\newcommand{\al} {{\alpha}}
\newcommand{\la} {{\lambda}}
\newcommand{\La} {{\Lambda}}

\newcommand{\ket}[1] {\left|#1\right>}

\newcommand{\vev}[1] {\left<#1\right>}

\title{Democratic Superstring Field Theory: Gauge Fixing}

\author{Michael Kroyter\\
Center for Theoretical Physics\\
Massachusetts Institute of Technology\\
Cambridge, MA 02139, USA\\
\\ and\\ \\
School of Physics and Astronomy\\
The Raymond and Beverly Sackler Faculty of Exact Sciences\\
Tel Aviv University, Ramat Aviv, 69978, Israel\\ \\
\email{mikroyt@mit.edu}, \email{mikroyt@tau.ac.il}
}

\abstract{
We show that a partial gauge fixing of the NS sector of the
democratic-picture superstring field theory leads to the non-polynomial
theory.
Moreover, by partially gauge fixing the Ramond sector we obtain
a non-polynomial fully RNS theory at pictures $0$ and $\frac{1}{2}$.
Within the democratic theory and in the
partially gauge fixed theory the equations of motion of both sectors
are derived from an action.
We also discuss a representation of the non-polynomial theory analogous to a
manifestly two-dimensional representation of WZW theory and the action of
bosonic pure-gauge solutions.

We further demonstrate that one can consistently gauge fix the NS sector of
the democratic theory at picture number $-1$. The resulting theory is new.
It is a $\Ztwo$ dual of the modified cubic theory. We construct analytical
solutions of this theory and show that they possess the desired properties.
}

\keywords{String Field Theory, Superstrings and Heterotic Strings}
\preprint{MIT-CTP-4171\\TAUP-2915-10}

\begin{document}

\section{Introduction}

Could string field theory serve as a non-perturbative definition of
superstring theory?
For that to be the case one needs a reliable formulation of superstring
field theory. Such a formulation should describe interactions of the
world-sheet modes in a consistent way. It should also respect the symmetries
of the theory, e.g., be
covariant. Moreover, since string theory can be defined around many
backgrounds, a good formulation should ``look the same'', regardless of
the background chosen. This requirement is not as strong as that of
``background independence'', since it allows the theory to depend
on the details of the background, e.g., on the BRST charge $Q$.
A ``universal formalism'' would depend only on ``universal objects'' such
as $Q$ that can be defined for any consistent background, regardless of
its dimensionality, specific matter content, etc.
Indeed, universality played a major role in the advance of string
field theory following Sen's
conjectures~\cite{Sen:1999mh,Sen:1999xm}\footnote{A review that covers recent
developments in the field, including some material that is relevant for this
work is~\cite{Fuchs:2008cc}. A shorter review of some of these developments
is~\cite{Schnabl:2010tb}.}.
Superstring field theory should also be able to describe the Ramond sector,
again, in a universal way. An open RNS string field theory, which we call the
``democratic theory'', that passes all the above criteria was constructed
in~\cite{Kroyter:2009rn}.

Of course, the above is only a preliminary list of the expected properties
of such a theory. First, the theory should support classical solutions
describing known physics. Such solutions exist in the democratic theory. Then,
for supersymmetric backgrounds, one might expect to obtain a supersymmetric string
field theory.
This was not fully established yet for the democratic theory. Note, however, that
supersymmetry is not a universal property. Moreover, the expectation that the full
string field theory is supersymmetric might be too naive.

Finally, it should be possible to
gauge fix the theory. Then, one should be able to identify propagators and
construct perturbation theory that leads to correct expressions for
scattering amplitudes.
The gauge structure of the democratic theory was identified and the classical
BV master action of it was constructed. However, the issue of gauge fixing
the theory was not clear. It is the purpose of this paper to clarify it to
some extent. Specifically, we show that different gauge choices for the NS
sector of the theory lead to the two known consistent NS string field
theories, namely the non-polynomial theory and the modified theory, as well
as to a new theory, which is the $\Ztwo$ dual of the modified theory
(the non-polynomial theory is self-dual).
Furthermore, the equations of motion of the Ramond sector of the democratic
theory seem to be equivalent to those proposed for the non-polynomial theory
and have the added advantage of being derivable from an action.

The rest of the paper is organized as follows: We begin by recalling the
definition of the democratic theory and setting our conventions in
section~\ref{sec:SSFTintro}.
Then, we start section~\ref{sec:gaugeToNP} with some algebraic preliminaries
related to the representation of the non-polynomial theory and to the
action of pure-gauge solutions and also present a simple two-dimensional
representation of the WZW theory.
We then show that the democratic theory can be
partially gauge fixed in a way that leads to the non-polynomial theory.
Moreover, we present a partially gauge fixed action generalizing the usual
non-polynomial one that includes the Ramond sector.
In section~\ref{sec:newSSFT} we derive a new theory that is obtained by gauge
fixing the NS sector of the democratic theory at picture number $-1$. Next, we
illustrate, in section~\ref{sec:solutions}, that this theory supports
analytical solutions of familiar form and that the properties of these
solutions agree with the expectations. Conclusions and future directions are
given in section~\ref{sec:conc}.

\section{Cubic superstring field theory in the democratic picture}
\label{sec:SSFTintro}

When a string field theory is constructed, one first identifies the vertex
operators of the string theory and generalizes them to an off-shell set of
string fields. Then, the equations of motion and gauge symmetry of the
theory are constructed with the restriction that their linearized versions
agree with the world-sheet conditions of being ``closed'' and ``exact''
respectively.

There are many ways for constructing a complete set of vertex operators of
the RNS string. The source of the complication is the existence of an
infinite number of pictures for each vertex~\cite{Friedan:1985ge}.
While the sets of vertex operators at different pictures are
equivalent~\cite{Horowitz:1988ip,Lian:1989cy}, their natural off-shell
generalizations are not. Hence, the structure of the resulting string field
theory might depend on the choice of picture.

A related subtlety is the existence of different ways for representing
vertex operators in the large Hilbert space
(henceforth, ``the large space'').
The space one gets from considering the ghost systems that are adequate
for dealing with the world-sheet symmetries of the RNS string is called the
small Hilbert space (henceforth, ``the small space''). The bosonization of
the $\beta\gamma$ superghost system
to the $\xi\eta\phi$ variables does not include the $\xi$ zero mode.
The large space is the space obtained by including this zero mode.
This space relates to the small space in a way that reminds the
relation between complex and real numbers:
\begin{itemize}
\item The large space is a natural extension of the small space, since the
      latter is obtained from the former by dropping a single mode of a
      conformal field.
\item It manifests the symmetries of the theory. Specifically, in the large
      space one can think of the RNS string as a topological $N=4$
      system~\cite{Berkovits:1994vy}. In this description, the BRST operator
      is dual to the $\eta$-ghost zero mode $\eta_0$. Both operators are
      nilpotent and they (anti-)commute. This $\Ztwo$ symmetry
      can be extended in some cases to act also on the spectrum of the
      theory~\cite{Berkovits:1996bf} (though not in a universal way).
\item The operators $Q$ and $\eta_0$ have trivial cohomologies in the large
      space. This means that every closed state has a primitive.
      Up to OPE singularities, these primitives can be obtained using the
      contracting homotopy operators of $Q$ and $\eta_0$, which are
      $\xi$ and $P\equiv -c \xi \xi' e^{-2\phi}$ respectively.
      While these primitives are ``unreal'' in the sense that they
      generically are not part of the small space (they are the
      analogues of $i\in \mathbb{C}$), they might enable the simplification
      of various constructions, e.g., that of picture changing operators.
\end{itemize}

The first attempt towards a universal RNS string field theory, by Witten,
was based on picture $-1$ string fields living in the small space.
This construction was shown to be inconsistent~\cite{Wendt:1987zh}.
Nonetheless, Berkovits' construction of a theory that is based on a large
space vertex operators at picture number zero (the immediate
analogues of the small space picture number $-1$ operators) led to a
consistent, non-polynomial theory~\cite{Berkovits:1995ab}.
An earlier attempt towards a theory was made by simply modifying the picture
number from $-1$ to
zero~\cite{Arefeva:1989cm,Arefeva:1989cp,Preitschopf:1989fc}.
This theory also seems to be consistent in the NS sector.

Despite the success of the ``non-polynomial'' and the ``modified'' theories,
one might wonder whether the zero picture is really what one should
ultimately use, especially since, as we already mentioned, the off-shell
structure is different at different pictures.
Other than this ``aesthetic'' question there exists also the problem of
including the Ramond sector: The formulation of the Ramond sector of the
modified theory is inconsistent~\cite{Kroyter:2009zi}, while in the
non-polynomial theory it is not known how to include it in a
universal way~\cite{Berkovits:2001im}, at least without ``tricks'' such as
imposing constraints after the derivation of the equations of
motion~\cite{Michishita:2004by}.

One can address these issues by building a theory upon
a different realisation of vertex operators. This realisation is obtained by
allowing the vertex operators to live in the large space and to have
an arbitrary picture number~\cite{Berkovits:2001us}, while modifying the
kinetic operator from $Q$ to $Q-\eta_0$. Equivalent vertex operators belong
to the same cohomology class regardless of their picture.
Since each cohomology class now lives in an infinite dimensional space,
inconsistencies might evolve, unless some restrictions are applied. A
natural such restriction is that the sum over pictures of the coefficients
of each vertex operator is well defined and finite.
With this restriction of the space of unbounded picture numbers,
it was shown in v2 of~\cite{Kroyter:2009rn}, that the cohomology
of $Q-\eta_0$ is the correct one,
i.e., it is identical to that of the same operator over an arbitrary
bounded range of picture operators.

A string field theory that generalizes this construction of vertex
operators was introduced in~\cite{Kroyter:2009rn}. Its string field $\Psi$
has ghost number one (as do the vertex operators), but an arbitrary picture
and it lives in the large space. String fields should be restricted
in a way that generalizes the restriction on vertex operators mentioned
above. However, since there is no canonical norm at our
disposal, it is not clear how exactly should this condition be quantified.
This is one more manifestation of the problem regarding the definition of
a proper space of string fields (see, e.g.,~\cite{Kroyter:2009zj}).
The action of this theory (henceforth ``the democratic theory'') is given
by\footnote{Here, we introduce the convention of the integration symbol for
a large space expectation value, as opposed to the standard integration
symbol which refers to either the small space or to the space of the bosonic
theory. Also, here and in the rest of the paper, we keep the star
product~\cite{Witten:1986cc} implicit. Another convention that we use is
that $[A,B]$ represents a graded (star) commutator.},
\begin{equation}
\label{action}
S=-\oint \cO \Big(\frac{1}{2}\Psi (Q-\eta_0)\Psi+\frac{1}{3}\Psi^3\Big),
\end{equation}
with $\cO$ being a mid-point insertion described below.

This theory addresses well the ``aesthetic challenge'', since it lives in the
large space and include the off-shell generalizations of all pictures.
Moreover, its BV master action takes the same form, only with the ghost
number of $\Psi$ being arbitrary. This means that the whole space of
NS string fields is used by the theory.
The second challenge, that of including the Ramond sector is resolved
by allowing $\Psi$ to be a linear combination of all integer picture
number NS fields and all half-integer number Ramond fields. We write,
\begin{equation}
\Psi=A+\al\,,
\end{equation}
where $A$ is the NS string field and $\al$ is the Ramond string field.
In terms of $A$ and $\al$ the action takes the form,
\begin{equation}
\label{RNSaction}
S= -\oint \cO\Big(\frac{1}{2}A \tQ A+\frac{1}{3}A^3
+\frac{1}{2}\al \tQ \al+A \al^2\Big),
\end{equation}
which is exactly what one would expect for an RNS theory.
Reverting to the aesthetic issue, we see that the theory uses all possible
component fields of the RNS theory and naturally unifies the NS and Ramond
component fields to a single entity $\Psi$. Moreover, the inclusion of
arbitrary D-brane systems can be straightforwardly achieved along the lines
of~\cite{Berkovits:2000hf} and the theory would then use the maximal set of
permissible component fields.

In order to complete the definition of the theory we have to specify the
mid-point insertion\footnote{One might worry that the use of a
mid-point insertion would lead to inconsistencies as in the cases of Witten's
theory and of the Ramond sector of the modified theory. This is not the case,
since the mid-point insertion appears only in the action. It is our
understanding that mid-point insertions should never appear on string fields,
since this would in any case lead to inconsistencies~\cite{Kroyter:2009zj}.
It is therefore important that the equations of motion and the gauge
symmetry of the theory do not
include mid-point insertions. This is indeed the case with the democratic
theory, in contrast with the situation of the inconsistent theories.} $\cO$.
The ``physical part'' of the vertex
operators that we generalize lives in the small space. Thus, $\cO$ should
include a component that is proportional to $\xi$.
In~\cite{Kroyter:2009rn}, we showed that gauge invariance of the action
together with $\xi$ as an ``initial condition'' implies that,
\begin{equation}
\label{cOfirstDef}
\cO\simeq \xi \sum_{k=-\infty}^\infty X_k\,,
\end{equation}
where $X_k$ is the picture changing operator of order $k$, e.g.,
\begin{equation}
X_0=1\,,\qquad X_1=X=Q\xi\,,\qquad X_{-1}=Y=\eta_0 P\,.
\end{equation}
The $\simeq$ symbol in~(\ref{cOfirstDef}) reminds us that the expression is
not really well defined and should be regularized in order to remove OPE
poles. It should further be modified in order to turn $\cO$ into a primary
conformal field. A more accurate definition of $\cO$ would be the statement
that it is a weight zero primary conformal field obeying,
\begin{equation}
\label{QOetaO}
Q\cO=\eta_0 \cO=\sum_{k=-\infty}^\infty X_k\,.
\end{equation}
One can also decompose this equality according to picture number and write,
\begin{equation}
\cO=\sum_{k=-\infty}^\infty \cO_k\,,\qquad Q \cO_k=X_k\,,\qquad
 \eta_0 \cO_k=X_{k-1}\,.
\end{equation}
It was shown in~\cite{Kroyter:2009rn} that such an $\cO$ exists and that
the resulting $X_k$ operators are also zero-weight primaries.
Furthermore, it was also shown there (following~\cite{Kroyter:2009bg}) that
different $\cO$ insertions obeying these relations differ by a $Q-\eta_0$
exact terms and they are all (at least) classically equivalent.

As already mentioned, the action~(\ref{action}) has a gauge symmetry,
which is a non-linear extension of the exactness property of vertex
operators. The infinitesimal form of the gauge transformation is,
\begin{equation}
\label{linGaugeSym}
\delta \Psi=(Q-\eta_0)\La+[\Psi,\La]\,,
\end{equation}
where $\La$ is the gauge string field\footnote{This terminology might be a
bit confusing, since the analogue of the QED gauge field $A_\mu$ is $\Psi$,
not $\La$. Due to this analogy one can find in the literature the term
``gauge parameter'' for $\La$. We find this terminology even more confusing,
since $\La$ is not a parameter, but a string field.}, which is even and
includes all integer (NS) and half-integer (R) picture components.
The zero-picture small-space component of $\La$ generates the standard
gauge transformations, while all other (NS) components generate picture
changing. Thus, ``most'' of the gauge symmetry is associated with
picture changing. The gauge symmetry~(\ref{linGaugeSym}) can be
exponentiated to give the finite form of a gauge transformation,
\begin{equation}
\label{gaugeTrans}
\Psi\rightarrow e^{-\La}\big(Q-\eta_0+\Psi\big)e^\La\,.
\end{equation}
Note, that the fact that we use an all-picture theory enabled us to write
a gauge symmetry with no explicit (mid-point) picture changing operators.
This feature is what enables the theory to be consistent, as opposed to
previous similar formulations.

The equations of motion and gauge symmetry immediately suggest a simple
way of gauge fixing in the NS sector. Requiring that the string field
and gauge string field both carry picture number zero and live in the small
space leads to the action,
\begin{equation}
S= -\int X_{-2}\Big(\frac{1}{2}A Q A+\frac{1}{3}A^3\Big),
\end{equation}
where $X_{-2}=Y_{-2}$ is the double inverse picture changing operator and
the remaining integration is in the small space. The residual gauge symmetry
is,
\begin{equation}
A\rightarrow e^{-\La}\big(Q+A\big)e^\La\,.
\end{equation}
We note that we obtained the modified theory. Indeed, this fact lied in the
heart of the construction of the democratic theory.
We would like to find other ways to partially gauge-fix
the theory, where ``partially'' refers to the picture-changing
gauge symmetry, but first, we would like to address some misconceptions
regarding this issue.

\subsection{Wrong ways to gauge fix}
\label{sec:GaugeFix}

A naive examination of the theory suggests that gauge fixing leads to
erroneous results. Specifically, one might claim that the gauge symmetry
related to picture number can be fixed to any single picture.
Then, it seems that choosing picture $-1$ for the NS field,
leads to Witten's theory, which is known to be inconsistent. Thus,
it seems that the democratic theory should also be inconsistent.
This observation is wrong, for at least three reasons:
\begin{enumerate}
\item It is not clear at all that fixing the gauge at any particular
      picture is a legitimate gauge choice.
\item Our theory is defined in the large space. In order to get to a
      theory such as Witten's, one has to make the further gauge choice
      of restricting the string field to the small space. This is
      in general an inconsistent choice.
\item Even if we could obtain Witten's theory, we would have not obtained
      the inconsistent gauge symmetry of this theory.
\end{enumerate}
We discuss these three points in the following three subsections.

\subsubsection{Gauge fixing to a given picture number}

The democratic theory mixes different picture numbers in the equations of
motion and in the gauge transformation. In the latter, generically all picture
numbers are intertwined. Furthermore, since the content of the (off-shell)
string field depends non-trivially on its picture, it seems unnatural to
assume that a single picture number can be used in order to fix the gauge
and the possibility to fix the gauge to one picture number does not imply that it
is possible to do that at any other picture number. Thus, the different
picture numbers should be examined on a case by case basis.
Given a picture number $n$ that can be used for a partial gauge fixing, one can
form other permissible partially gauge fixed sets by considering an arbitrary
gauge string field $\La$ and defining the set,
\begin{equation}
\label{HLa}
H_\La\equiv\{A:\ \exists \hat A,\ pic(\hat A=n),\ A=e^{-\La}(Q-\eta_0+\hat A)e^\La\}\,.
\end{equation}
These sets generically include all picture numbers. Thus, gauge fixing to a
given picture number does not seem, a-priori, to be equivalent to gauge fixings
to other picture numbers.

In order to rule out most picture numbers from the list of permissible
partial gauge fixings, we note the following.
String fields are not allowed to carry mid-point insertions.
Hence, when we look at the equations of motion, we should actually demand
that the coefficient of each $\cO_k$ separately vanishes. Then, if we set
the picture numbers to arbitrary values, the linear terms and interaction
terms decouple and we are left with,
\begin{equation}
QA=\eta_0 A=A^2=\al^2=\ldots=0\,.
\end{equation}
While these equations are not wrong by themselves, they are certainly too
restrictive to allow for general solutions of the equations of motion.
Hence, they cannot be part of a permissible gauge fixing, since, by definition,
a gauge fixing should leave us with a representative of each
gauge orbit~\cite{Henneaux:1992ig}\footnote{In the case of a full gauge fixing
this representative should also be unique. For partial gauge fixing, as we
consider here, there should actually be many representatives of each gauge orbit.}.
The conclusion of this discussion is that, in general, a permissible gauge
fixing is not attained by fixing the picture number to some given value.

In order to further understand this issue, we note that fixing the picture number
of the string field implies also an appropriate constraining of the gauge string field.
Then, in order not to generate all possible pictures, the gauge string field should
carry zero picture number. The only contributions to the string field come now
at pictures zero and $-1$. This fact turns those picture
numbers into the only permissible ones for a fixed picture gauge choice in
the NS sector.
Furthermore, this state of affairs complicates the gauge fixing
of the Ramond sector, which carries a half-integer picture.

\subsubsection{Reducing the large space to the small (or dual) space}

From the point of view of vertex operators, the restriction to the small
space using $\eta_0$ and the condition of closeness that is imposed by $Q$,
stand on the same footing. The $\Ztwo$ symmetry between $Q$ and $\eta_0$
is manifest when we write the closeness condition and exactness relation as,
\begin{equation}
Q V=\eta_0 V=0\,,\qquad V\sim V' \ \Leftrightarrow\ V-V'=Q\eta_0 \Upsilon\,.
\end{equation}
In fact, one can interpret this relations not as saying
\begin{equation}
Q V=0\,,\qquad V\sim V' \ \Leftrightarrow\ V-V'=Q \La\,,
\end{equation}
with $V$ and $\La$ in the small space, but as saying that,
\begin{equation}
\eta_0 V=0\,,\qquad V\sim V' \ \Leftrightarrow\ V-V'=\eta_0 \La\,,
\end{equation}
where $V$ and $\La$ are restricted to live in the space that is obtained by
acting on the large space with $Q$. This space is the $\Ztwo$ dual of the
small space and we refer to it as ``the dual space''.
The fact that one should decompose the equations of motion according to the
picture number ($\cO$ components), might lead to a situation where the choice
of the small space is inconsistent. Indeed, we will see
that a simple restriction to picture number $-1$ leads not to Witten's
theory, but to a new theory that lives in the dual space and is the $\Ztwo$
dual of the modified theory.

\subsubsection{Gauge symmetry in Witten's theory}

Finally, we want to understand the nature of the inconsistency of Witten's
theory.
In this theory singularities appear in the gauge symmetry, as well
as in the evaluation of scattering amplitudes. It might be the case, that the
latter ones originate from a bad gauge fixing (e.g., the Siegel gauge might
be inadequate in this case).
Another problem with Witten's theory is the absence of a tachyon vacuum,
as can be seen using low-level level
truncation~\cite{DeSmet:2000je}\footnote{Contrary to what was
written in previous versions of this paper, the analysis
of~\cite{DeSmet:2000je} is reliable and the choice of internal
Chan-Paton factors there is effectively identical to that
of~\cite{Arefeva:2002mb}. I am grateful to the referee for
clarifying to me the conventions of~\cite{DeSmet:2000je}.

That Witten's theory cannot support a tachyon vacuum can be proven
without referring at all to the level-truncation.
The reason for that is that, according to our understanding of the role
of mid-point insertions, the equations of motion
are $Q A=A^2=0$. Hence, the action of all solutions must equal zero and
tachyon solutions cannot exist (on the other hand, solutions describing
marginal deformations might exist).}.
The problem with Witten's theory can be traced to the absence of
an analogue of the state $c(0)\ket{0}$ in the $-1$
picture~\cite{Aref'eva:2000mb}. This state cannot be picture-changed to the
$-1$ picture, since it is not on-shell and an attempt to ``carry it over
anyway'' using $Y$ leads to zero. The importance of this state is that it is
an auxiliary field, which when integrated out gives a quartic potential for
the tachyon field.

At any rate, the problem with the gauge symmetry of Witten's theory is
genuine. This gauge symmetry includes mid-point insertions of picture
changing operators that collide upon iteration and lead to singularities.
Even if we would have gotten this theory as a result of a partial gauge
fixing of the
democratic one, we would have not gotten this gauge symmetry, since we do not
allow mid-point insertions over string fields. Hence, we would have concluded
that the theory has no gauge symmetry at all. Such a theory should have been
also considered as inconsistent, since it would not lead to a correct space
of vertex operators upon linearization. However, we do not get this theory by
gauge fixing the democratic one. We study the gauge fixing of the democratic
theory to picture number $-1$ in section~\ref{sec:newSSFT}, but first we turn
to another gauge fixing, which leads to the non-polynomial theory.

\section{Gauge fixing to the non-polynomial theory}
\label{sec:gaugeToNP}

We now want to illustrated that the NS sector of the democratic theory
reduces to the non-polynomial theory upon a partial gauge fixing.
We start by presenting some useful algebraic structures in~\ref{sec:Prel},
where we also discuss the action of bosonic pure gauge solutions.
The main construction follows in~\ref{sec:GaugeFixingToNP}. We then discuss
the gauge fixing of the Ramond sector in~\ref{sec:Ramond}.

\subsection{Preliminaries}
\label{sec:Prel}

Following~\cite{Gates:1983nr}, we define the operator $L$ as,
\begin{equation}
\label{Ldef}
L A=[\Phi,A]\,. 
\end{equation}
Such an operator can be defined for an arbitrary $\Phi$ and thus could be
better named $L_\Phi$. We would mostly be using this operator for whatever
string field that we call ``$\Phi$''. Hence, in this case, we avoid the
subscript.
The operator $L$ is an even derivation. This implies the following
integration by parts formula (the same holds for integration in the large
space),
\begin{equation}
\label{intByParts}
\int f(L)A_1 A_2= \int A_1 f(-L) A_2\,.
\end{equation}

One can formally represent the derivations $Q$ and $\eta_0$ as
commutators~\cite{Horowitz:1986dt},
\begin{equation}
\label{derAsComm}
Q A=[A_Q,A]\,,\qquad \eta_0 A=[A_\eta,A]\,,
\end{equation}
where we define the formal states $A_Q$ and $A_\eta$ as line integrals over
the identity state. In the cylinder coordinates they take the form,
\begin{subequations}
\label{AQAeta}
\begin{align}
A_Q &=-\int_{-i\infty}^{i\infty} J_B (z)dz \ket{\One},\\
A_\eta &=-\int_{-i\infty}^{i\infty} \eta (z)dz \ket{\One}.
\end{align}
\end{subequations}
We can also write,
\begin{equation}
Q\Phi=-L A_Q\,,\qquad \eta_0\Phi=-L A_\eta\,.
\end{equation}

The action of the non-polynomial theory can be cast into the
form\footnote{This result was obtained in the course of a work in progress
on the gauge structure of the non-polynomial theory, carried out in
collaboration with Nathan Berkovits, Yuji Okawa, Martin Schnabl, Shingo
Torii and Barton Zwiebach. The starting point for this derivation was the
representation of the non-polynomial theory found by Berkovits, Okawa and
Zwiebach in~\cite{Berkovits:2004xh}.},
\begin{equation}
\label{S2D}
S_{NP}=-\oint \eta_0\Phi \frac{e^{-L}-1+L}{L^2}Q\Phi\,.
\end{equation}
By expanding the exponent and using~(\ref{Ldef}), this expression is seen
to be equal to,
\begin{equation}
S_{NP}=-\oint \eta_0\Phi \Big(\frac{1}{2}Q\Phi-\frac{1}{6}[\Phi,Q\Phi]
  +\frac{1}{24}\big(\Phi^2 Q\Phi-2\Phi Q\Phi\Phi+Q\Phi \Phi^2\big)+\ldots\Big),
\end{equation}
which agrees with eq.~(2.35) of~\cite{Berkovits:2000hf}
(up to a global factor of 2 stemming from the change of normalization due to
the inclusion of the GSO($-$) sector).

Using~(\ref{Ldef}) and~(\ref{derAsComm}), the action~(\ref{S2D}) can be
rewritten in the more suggestive form,
\begin{equation}
\begin{aligned}
\label{SNP2D}
S_{NP} &=-\oint [A_\eta,\Phi] \frac{e^{-L}-1+L}{L^2}[A_Q,\Phi]=
 -\oint L A_\eta \frac{e^{-L}-1+L}{L^2}L A_Q\\&=
\oint A_\eta (e^{-L}-1+L) A_Q=
 \oint A_\eta e^{-L} A_Q\,.
\end{aligned}
\end{equation}
Here, we used~(\ref{intByParts}) in the second equality and in the last
equality we dropped one term in light of the fact that both integrands of
$A_Q$ and $A_\eta$ live in the small space~(\ref{AQAeta}) and we
dropped another term using,
\begin{equation}
[A_Q,A_\eta]=0\,.
\end{equation}

Another way to prove that the action~(\ref{SNP2D}) equals the standard non-polynomial
one is by considering the variation of both actions. Specifically, given $\Phi$,
we can consider the one-parameter family connecting it to the trivial configuration,
$\Phi(\al)\equiv \al \Phi$ $(0\leq \al\leq 1)$. The action of $\Phi(\al)$ is
obtained by replacing $L$ in~(\ref{SNP2D}) by $\al L$.
Evaluating the derivative with respect to $\al$ leads to,
\begin{equation}
\label{NPvar}
\frac{dS_{NP}}{d\al} = \frac{d}{d\al}\oint A_\eta e^{-\al L} A_Q
  =-\oint A_\eta L A=\oint L A_\eta A=-\oint \eta_0\Phi A
  =\oint \eta_0 A \Phi \,.
\end{equation}
The same expression is obtained from the familiar variation of the standard
representation of the non-polynomial action
\begin{equation}
\delta S_{NP}	= \oint \eta_0 A e^{-\Phi}\delta e^\Phi\,,
\end{equation}
upon replacing the variation by a derivative with respect to $\al$.
Since both actions obey the same first order differential equation with the
same (trivial) initial conditions they must agree.

The representations~(\ref{S2D}) and~(\ref{SNP2D}) are analogous to
manifestly two dimensional representations of WZW theory. The former
appears to be less attractive, but can be written directly for the WZW case.
All that is needed is to replace the $Q$ and $\eta_0$ by the derivatives
$\partial$ and $\bar \partial$, to reinterpret the string field as an algebra
element in WZW theory (also reinterpret $L$ accordingly) and to replace
the integration by a combination of complex plane integration and a trace
over group space. The representation~(\ref{SNP2D}), which looks simpler,
includes the string fields $A_Q$ and $A_\eta$.
Their counterparts in the WZW case are the derivatives
$\partial$ and $\bar \partial$, considered as operators, i.e., acting
all the way to their right (and further on, when we take the cyclicity
property of the trace into account).
A manifestly two-dimensional representation of the WZW theory might be
counter-intuitive, since it is known that the coupling constant should be
quantized and one might worry that we do not get this condition in the
current representation.
This is not the case. Given $g$ (the analogue of $A$ in the WZW case)
a choice of $\Phi$ implies a choice of cohomology class. In order to obtain
an action that is independent on the latter choice (up to the addition of
$2\pi n$), the usual quantization condition~\cite{Witten:1983ar}
should hold in our new representation. For non-compact groups, as usual,
no quantization would arise, since the relevant cohomology class is trivial.
This is similar to the string field theory case.

It is not known how to extend the non-polynomial theory to the Ramond sector
in a universal way. What is known~\cite{Michishita:2004by} is a way to write
an action with two Ramond string fields and a constraint that should be
imposed only after the derivation of the equations of motion.
The action with the two Ramond string fields $\Xi$ and $\Psi$ can be cast
into the form,
\begin{equation}
S_{NP} =\oint \hat A_\eta e^{-L} \hat A_Q\,,
\end{equation}
where we defined,
\begin{equation}
\hat A_\eta \equiv A_\eta+ \frac{[A_\eta,\Psi]}{\sqrt{2}}\,,\qquad
\hat A_Q \equiv A_Q+ \frac{[A_Q,\Xi]}{\sqrt{2}}\,,
\end{equation}
i.e., the operators that generate the derivations should be modified in
order to include the influence of the Ramond string field. The factors
of $\sqrt{2}$ originate from the fact that we use two Ramond string fields.
This is similar to the construction in section 5 of~\cite{Kroyter:2009zi}.
For completeness we recall that the constraint relating the picture
$\frac{1}{2}$ string field $\Psi$ and the picture $-\frac{1}{2}$ string
field $\Xi$ is,
\begin{equation}
\eta_0 \Psi=e^{-L} Q\Xi\,.
\end{equation}

\subsubsection{Example of using $A_Q$: The action of bosonic pure-gauge solutions}
\label{sec:DirectGaugeinv}

It is a ``common knowledge'' that one cannot prove that the action vanishes
for pure-gauge bosonic solutions, in terms of the finite form of the
gauge transformation. Specifically, one has to prove that,
\begin{equation}
\int e^{-\La}Q e^\La Q e^{-\La} Q e^\La =0\,,
\end{equation}
which cannot be achieved just by integrating by parts and using the trace
property.
However, if one writes the infinitesimal form of the gauge transformation,
\begin{equation}
\label{cQ}
\delta A=Q\La+[A,\La]\equiv \cQ \La\,,
\end{equation}
and uses the fact that $A$ is a solution, i.e., that the action equals,
\begin{equation}
S=-\frac{1}{6}\int A Q A\,,
\end{equation}
one gets
\begin{equation}
\delta S=-\frac{1}{6}\int \big(\cQ \La Q A+A Q\cQ \La)=0\,.
\end{equation}
Then, the fact that the action of a pure-gauge solution is zero can be
established, since $S$ obeys the first order linear differential equation
and initial condition,
\begin{equation}
\frac{dS}{d \al}=0\,,\qquad S(0)=0\,,
\end{equation}
where we defined,
\begin{equation}
\label{paramPureGauge}
A(\al)=e^{-\al \La} Q e^{\al \La}\,.
\end{equation}
One usually says that the proof holds provided it is allowed to use the
parametrization~(\ref{paramPureGauge}). This condition is what distinguishes
a genuine gauge solution from a formal one.

Let us now write the pure gauge solution in terms of the formal string field
$A_Q$,
\begin{equation}
A=e^{-\La}Q e^\La=\big(e^{-L_\La}-1\big)A_Q\,.
\end{equation}
Using the trace property we can write,
\begin{equation}
\label{AQcube}
S=\frac{1}{6}\int A^3=\frac{1}{6}\int \Bigg(\big(e^{-L_\La}A_Q\big)^3
 +3\big(e^{-L_\La}A_Q\big)^2 A_Q +3e^{-L_\La}A_Q A_Q^2 + A_Q^3\Bigg).
\end{equation}
The last two terms vanish in light of the identity,
\begin{equation}
\label{AQ20}
A_Q^2=0\,.
\end{equation}
The fact that $L_\La$ is a derivation implies that for arbitrary $A_n$
the following holds\footnote{The operator $e^{L_\La}$ generates the
``Taylor expansion'' for the derivation $L_\La$.},
\begin{equation}
e^{L_\La} \big(A_1\ldots A_n\big)
  =e^{L_\La} A_1\ldots e^{L_\La} A_n\,.
\end{equation}
This identity together with~(\ref{AQ20}) imply that the first two terms
in~(\ref{AQcube}) drop out as well. Thus, we managed to prove that the action
of a pure-gauge solution vanishes using the finite form of the gauge
transformation.
One can now define a formal gauge string field as one for which any of the
manipulations/representations used in our construction do not hold.

\subsection{The gauge fixing}
\label{sec:GaugeFixingToNP}

Let us examine first the case of the linearized theory of the NS sector,
\begin{equation}
S= - \frac{1}{2} \oint \cO A (Q-\eta_0) A\,.
\end{equation}
We want to partially fix the gauge by requiring $A$ to have picture number
zero,
\begin{equation}
\label{pic0}
pic(A)=0\,.
\end{equation}
The equations of motion then reduce to,
\begin{equation}
QA=\eta_0 A=0\,.
\end{equation}
The gauge symmetry at this level reads\footnote{This form of the gauge
symmetry does not use the linearized equations of motion. Instead, it uses
the gauge condition~(\ref{pic0}). See also the discussion
in~\ref{sec:ResGauge}.},
\begin{equation}
\delta A=Q \eta_0 \Upsilon\,.
\end{equation}
We can eliminate the gauge symmetry by restricting the string field not to be
part of either the small space or the dual space. The two options are
equivalent and we find it convenient to choose the second and define,
\begin{equation}
\label{AQPhi}
A=Q\Phi\,.
\end{equation}
Equations~(\ref{pic0}) and~(\ref{AQPhi}) constitute our gauge conditions.
Note, that by introducing $\Phi$ as a solution to the gauge condition
$Q A=0$, we introduce an extra gauge symmetry, namely
\begin{equation}
\Phi \rightarrow \Phi + Q\hat \La\,.
\end{equation}

With our gauge conditions the action can be written as,
\begin{equation}
S= \frac{1}{2} \oint \cO Q\Phi \eta_0 Q \Phi=
  \frac{1}{2} \oint (Q\cO) Q\Phi \eta_0 \Phi=
  -\frac{1}{2} \oint \eta_0 \Phi Q\Phi \,.
\end{equation}
Here, we integrated by parts in the second equality and used~(\ref{QOetaO})
and the picture number of the integrand in order to set $Q\cO$ to unity in
the last equality. We recognize the final result as the linearized action
of the non-polynomial theory.

Consider now the interacting NS theory. As our gauge condition we take
again~(\ref{pic0}) and for the second condition we choose what might be
considered as the natural non-linear generalization of~(\ref{AQPhi}), namely
\begin{equation}
\label{AofPhi}
A=e^{-\Phi}Q e^\Phi=\big(e^{-L}-1\big) A_Q\,.
\end{equation}
Note that this is {\it not} (even a formal) pure-gauge form, since the
kinetic operator of the democratic theory is $Q-\eta_0$ and not $Q$.
In particular, the resulting $A$ does not necessarily obey the equation of
motion. Also, we do not restrict $\Phi$ to obey the equation of motion of the
non-polynomial theory. While such a requirement was necessary in previous
constructions of this type, this is not necessary here, since the string
field $A$ lives in the large space.
Again, writing $A$ in terms of $\Phi$ introduces an extra gauge symmetry.
This gauge symmetry is generated by,
\begin{equation}
\label{La0GaugeSym}
\delta e^\Phi = Q\hat \La_0 e^\Phi\,,
\end{equation}
with $\hat \La_0$ being an arbitrary picture zero odd string field.

We want to prove that this gauge fixing leads to the non-polynomial theory.
To that end, we first note, in~\ref{sec:EOM}, that the equation of motion is
obeyed simultaneously in both theories. Next, in~\ref{sec:ResGauge}, we
prove that there is a one to one correspondence between the gauge orbits
of the non-polynomial theory and the residual gauge orbits of the democratic
theory. Finally, in~\ref{sec:Action}, we prove that the action coincides for
both theories {\it even off-shell}. All that proves that the non-polynomial
theory is a partially gauge-fixed version of the (NS sector of the)
democratic theory (assuming that our gauge choice is a legitimate one).

Note, that one can choose gauge transformed versions of
our gauge condition~(\ref{AofPhi}). Substituting $\La=-\Phi$
into~(\ref{gaugeTrans}), leads to,
\begin{equation}
\label{AofEta}
A=-e^\Phi\eta_0 e^{-\Phi}\,,
\end{equation}
while substituting $\La=-\frac{\Phi}{2}$ leads to the symmetrized expression,
\begin{equation}
\label{AofQeta}
A=e^{-\frac{\Phi}{2}}Q e^{\frac{\Phi}{2}}
  -e^{\frac{\Phi}{2}}\eta_0 e^{-\frac{\Phi}{2}}\,.
\end{equation}
These variants have different pictures, but they are gauge equivalent.
Hence, in light of the proof mentioned above, they are all equivalent to the
non-polynomial theory.
Their existence manifests the $\Ztwo$ symmetry of both theories.
We now turn to the actual proof.

\subsubsection{Equations of motion}
\label{sec:EOM}

The constraint~(\ref{pic0}) reduces the equations of motion to,
\begin{equation}
QA+A^2=\eta_0 A=0\,.
\end{equation}
The first of this equations is automatically obeyed in light
of~(\ref{AofPhi}). Indeed, it seems to us that a consistent gauge fixing of
the freedom related to the use of the large space could only be achieved by
choosing a gauge that enforces one of these two equations (or a linear
combination thereof). Thus, neglecting the possibility of enforcing a linear
combination of the equations, a consistent
gauge fixing to the zero picture would either lead to the non-polynomial
theory, as we do here, or to the modified theory, should the second
equation be chosen as the gauge condition.

With our choice of the gauge fixing, the second equation takes the form,
\begin{equation}
\eta_0\big(e^{-\Phi}Q e^\Phi\big)=0\,,
\end{equation}
which is exactly the equation of motion of the non-polynomial theory. Hence,
the equations of motion hold in one theory if and only if they hold in the
other.

\subsubsection{Residual gauge symmetry}
\label{sec:ResGauge}

Consider the infinitesimal form of the gauge transformations.
This is completely general, since the string fields
are not restricted to be infinitesimal.
The non-polynomial theory has two types of gauge symmetry,
\begin{equation}
\label{NPgaugeTrans}
\delta e^\Phi=Q\hat \La_0 e^\Phi+e^\Phi \eta_0 \hat \La_1\,,
\end{equation}
with the subscript referring to the picture number and the two gauge string
fields are odd. We use the hat in order to distinguish these generators from
those of the gauge string fields of the democratic theory.
We already noticed that the first of these two transformations is just a new
gauge symmetry that is induced by our parametrization~(\ref{AofPhi}).
The second transformation does induce a change of $A$.
Note, that had we chosen~(\ref{AofEta}) or~(\ref{AofQeta}) for our $A$,
the roles of the two types of gauge transformation would have changed, but
the final conclusion would have not.
The change of $A$~(\ref{AofPhi}) that is induced
by a general change of $\Phi$ is,
\begin{equation}
\label{deltaA}
\delta A=e^{-\Phi}\big(Q\delta e^\Phi-\delta e^\Phi A\big)\,.
\end{equation}
Plugging~(\ref{NPgaugeTrans}) into~(\ref{deltaA}) leads to,
\begin{equation}
\label{AcQetaLa}
\delta A=\cQ \eta_0 \hat \La_1\,,
\end{equation}
where $\cQ$ is defined as in~(\ref{cQ}).

In the case of the democratic theory,
restricting only to transformations that do not alter~(\ref{pic0}), leads to,
\begin{equation}
\label{deltaAgaugeFixed}
\delta A=\cQ \La_0-\eta_0 \La_1\,,  
\end{equation}
where the gauge string fields $\La_k$ are restricted to obey,
\begin{equation}
\label{cQLak}
\cQ \La_k=\eta_0 \La_{k+1}\,, \qquad -\infty<k<\infty\,,\ k\neq 0\,.
\end{equation}
While the operator $\cQ$ is often named ``the cohomology operator
around the solution $A$'', we remind that in the case at hand $A$ is
generally not a solution. On the other hand, had we been thinking of $A$ as
a solution of a theory whose kinetic operator is $Q$, it would have been a
pure-gauge solution.
The variation $\delta A$ should be further restricted in order to
maintain~(\ref{AofPhi}).
A necessary condition for that is
\begin{equation}
\label{deltaCQA}
\delta \big(Q A+A^2\big)=\cQ \delta A=0\,.
\end{equation}
From the definition~(\ref{cQ}) and the gauge choice~(\ref{AofPhi})
it follows that,
\begin{equation}
\label{cQsatisfy}
\cQ^2=0\,.
\end{equation}
Hence, the general gauge transformation induced by the non-polynomial
theory~(\ref{AcQetaLa}) obeys~(\ref{deltaCQA}). In fact, the
simplest resolution of~(\ref{deltaAgaugeFixed}) and~(\ref{deltaCQA}) is to
set
\begin{equation}
\label{gaugeForGaugeCond}
\La_k=0\qquad \forall k\neq 0\,.
\end{equation}
The remaining condition of~(\ref{cQLak}) is then,
\begin{equation}
\eta_0 \La_0 =0\,,
\end{equation}
which can be immediately solved to give~(\ref{AcQetaLa}), provided we
identify,
\begin{equation}
\La_0=\eta_0 \hat \La_1\,.
\end{equation}

The condition~(\ref{gaugeForGaugeCond}) can be thought of as a gauge
condition for the gauge string fields. This is sensible, since the gauge
symmetry of the theory is reducible and includes ``a gauge for gauge
symmetry'', ``a gauge for gauge for gauge symmetry'',  and so on, ad
infinitum. However, the gauge for gauge symmetry of the theory uses the
equations of motion. Hence, it might be the case that the condition can
be justified only on-shell. This should be enough for our purposes, since
``Frobenius' theorem'' implies that gauge orbits are generally well defined
only on-shell~\cite{Henneaux:1989jq,Henneaux:1992ig,Gomis:1994he}\footnote{In
our case gauge orbits can be defined off-shell, but the gauge for gauge
orbits cannot be defined, since these gauge symmetries do use the equations
of motion.}.
Imposing the equations of motion leads to,
\begin{equation}
[\eta_0,\cQ]=0\,.
\end{equation}
The operator $\cQ$ is trivial in the large space since,
\begin{equation}
\cQ B=0\quad \Rightarrow\quad Q e^L B=0\quad \Rightarrow\quad
 e^L B=Q\Upsilon\quad \Rightarrow\quad B=\cQ e^{-L} \Upsilon\,.
\end{equation}
These two facts are enough to enable pushing the $\La_k$'s all the way
to the gauge choice~(\ref{gaugeForGaugeCond}), in a similar manner to
the proof of the equivalence of $Q$ in the small space at a given picture
and $Q-\eta_0$ in the large space with arbitrary picture (assuming a
bounded or decaying behaviour as a function of picture
number)~\cite{Berkovits:2001us}. We conclude that the gauge orbits associated
with the residual gauge symmetry agree with those of the non-polynomial
theory.

\subsubsection{The action}
\label{sec:Action}

Using~(\ref{AofPhi}), the partially gauge fixed action of the democratic
theory can be written as,
\begin{equation}
\label{quadAction}
S=-\oint \cO\Big(\frac{1}{6}A Q A-\frac{1}{2}A \eta_0 A\Big).
\end{equation}
Note, that for this form of the action the allowed variation for $A$ is not
the most general one, due to the constraint~(\ref{deltaCQA}). Thus, the
generalization of the fundamental lemma of the calculus of variations does
not hold and the action cannot be used for deriving the equation of motion.
The way out is either to add to the action Lagrangian multipliers enforcing the constraint
or to solve the constraint and rewrite the action in
terms of $\Phi$. The former option cannot be achieved covariantly
without generating new unphysical degrees of freedom for the multiplier.
The latter option, on the other hand, leads exactly to the non-polynomial
action as we now prove.

One could have hoped to use the formal string fields $A_Q$ and $A_\eta$ in
order to prove the equality of the actions. However, the presence of the
$\cO$ mid-point insertion complicates the story, as it induces poles that
should somehow be regularized. Ignoring this issue would lead to erroneous
results. From the example in~\ref{sec:DirectGaugeinv}, we can infer that
if we are to avoid the use of $A_Q$ and $A_\eta$, we should not hope to be
able to derive the results with the finite form of $A$. However, the same
example also suggests a way out. We should write,
\begin{equation}
\label{Aal}
A(\al)=e^{-\al \Phi}Q e^{\al \Phi}\,,
\end{equation}
and prove that $\frac{dS}{d\al}$ obtains the same value in both theories.
The equality of the initial condition is trivial, since in both cases the
string field at $\al=0$ vanishes and so does the action.

We first note that,
\begin{equation}
\frac{d}{d\al} A=\cQ \Phi\,.
\end{equation}
Explicitly deriving the first term in the action with respect to $\al$ leads to,
\begin{equation}
\frac{d}{d\al}\Big(-\frac{1}{6}\oint \cO A Q A\Big)=
\frac{d}{d\al}\Big(\frac{1}{6}\oint \cO A^3\Big)=
\frac{1}{2}\oint \cO A^2 \cQ \Phi=
-\frac{1}{2}\oint \cO Q A \cQ \Phi\,.
\end{equation}
Substituting we get,
\begin{equation}
\begin{aligned}
\label{dSdal}
\frac{dS}{d\al} &=\frac{1}{2}\oint \cO \Big(\cQ \Phi \eta_0 A
 +A \eta_0\cQ\Phi-Q A\cQ \Phi\Big) \\ &=
\frac{1}{2}\oint \cO \Big(Q \Phi \eta_0 A
 +A \eta_0 Q\Phi-Q A Q \Phi + 2Q A \eta_0 \Phi\Big),
\end{aligned}
\end{equation}
where in the second line we have rewritten $\cQ$ in terms of $Q$ and used
\begin{equation}
\oint \cO A [B,C]=\oint \cO [A,B]C\,.
\end{equation}
Integrating by parts the first and third terms in~(\ref{dSdal}) gives,
\begin{equation}
\label{dSdalFin}
\frac{dS}{d\al} =\oint \cO \big(Q A \eta_0 \Phi-A Q\eta_0 \Phi\big)=
  \oint (Q\cO) A \eta_0 \Phi\,.
\end{equation}
We know from~(\ref{QOetaO}) that $Q\cO$ is just a sum of picture changing
operators. 
Inspecting the rest of the integrand reveals that the only one that
contributes is $X_0=1$.
Hence we obtain,
\begin{equation}
\label{dSdalFin1}
\frac{dS}{d\al} = \oint A \eta_0 \Phi\,,
\end{equation}
in agreement with the expression obtained for the non-polynomial theory~(\ref{NPvar}).
We conclude that the action
is the same for both theories {\it even off-shell}.
We thus finished establishing that the non-polynomial theory is a
partially gauge fixed version of the democratic one.

\subsection{The Ramond sector}
\label{sec:Ramond}

The equations of motion for the Ramond sector of the non-polynomial theory
were established by Berkovits in~\cite{Berkovits:2001im}, where it was
claimed that they cannot be derived from an action.
However, it was explicitly assumed there
that the action does not include any mid-point insertions.
We would like to show that we can get these equations of motion by a
partial gauge fixing of the democratic theory. We repeat the analysis twice.
First, in~\ref{sec:Direct}, we compare the equations of motion and
on-shell gauge symmetries of both sides. Then, in~\ref{sec:NP_RNSaction}, we
explicitly write down the partially gauge (picture) fixed action, which is a
non-polynomial, fully RNS action, with string fields at pictures 0 and
$\frac{1}{2}$.

\subsubsection{Direct comparison}
\label{sec:Direct}

Before attempting to compare the two theories, we have to settle some
conventions. As presented here, the democratic
theory has $Q-\eta_0$ as its kinetic operator. The conventions
of~\cite{Berkovits:2001im}, on the other hand, are adequate for the kinetic
operator $Q+\eta_0$. It is easy to see that a theory based on this operator
is dual to the one we are using here. The transformations between the two
are established by appending a minus sign to all the NS string fields, whose
picture number is odd as well as to all even picture number mid-point
insertions.
For the Ramond string fields it is immaterial which ones get the minus sign,
as long as the signs alternate as a function of the picture number.
To summarize, one can choose,
\begin{equation}
\label{etaMinusEta}
\Psi_k\rightarrow (-1)^{\lfloor k \rfloor} \Psi_k\,,\qquad
 \cO_k\rightarrow -(-1)^k \cO_k\,, \qquad \eta_0 \rightarrow -\eta_0\,.
\end{equation}

With our conventions for the kinetic operator and with~(\ref{etaMinusEta})
in mind, the equations of motion of~\cite{Berkovits:2001im} take the form,
\begin{equation}
\eta_0\big(e^{-\Phi}Q e^\Phi\big)-\big(\eta_0 \Xi\big)^2=0\,,\qquad
 \cQ \big(\eta_0 \Xi\big)=0\,,
\end{equation}
where $\Xi$ is a picture $\frac{1}{2}$ even Ramond string field.
The fact that these equations include only $\eta_0 \Xi$ implies the
existence of a gauge symmetry associated with this string field, which can
be solved by defining,
\begin{equation}
\label{alOfXi}
\al=\eta_0 \Xi\,.
\end{equation}
Now, $\al$ is the (odd) Ramond string field. It carries picture number
$-\frac{1}{2}$,
\begin{subequations}
\label{alGaugeCond}
\begin{equation}
pic(\al)=-\frac{1}{2}\,,
\end{equation}
and is constrained to obey,
\begin{equation}
\label{etaAl0}
\eta_0 \al=0\,.
\end{equation}
\end{subequations}
The equations of motion are,
\begin{equation}
\label{EOM1Rfield}
\eta_0\big(e^{-\Phi}Q e^\Phi\big)-\al^2=0\,,\qquad \cQ \al=0\,.
\end{equation}

Consider now the democratic theory and restrict the NS fields, as before, to
obey~(\ref{pic0}) and~(\ref{AofPhi}). For the Ramond string field we choose
the gauge conditions~(\ref{alGaugeCond}). Decomposing the equations of motion
according to picture number leads exactly to~(\ref{EOM1Rfield}). This is
hardly a surprise, since it was shown already in~\cite{Berkovits:2001im}
that the sum of the Ramond and NS string fields obey an equation of motion
with $Q+\eta_0$ as the kinetic operator. What is new, is the realisation
that these equations of motion {\it can} be derived from an action.

The gauge symmetry of~(\ref{EOM1Rfield}), identified
in~\cite{Berkovits:2001im}, can be rewritten as,
\begin{subequations}
\label{NPRgaugeof1R}
\begin{align}
\delta A=& \cQ \eta_0\hat\La_1+[\al,\cQ\hat\La_{\frac{1}{2}}]\,,\\
\label{gaugeOfAlOneR}
\delta \al=& [\al,\eta_0\hat \La_1]-\eta_0 \cQ \hat \La_{\frac{1}{2}}\,.
\end{align}
\end{subequations}
We already saw that the residual NS gauge symmetry of $A$ takes just this
form in the democratic theory. In our case, this gauge symmetry produces also
a picture $-\frac{1}{2}$ piece, which takes exactly the form of the first
term in the r.h.s of~(\ref{gaugeOfAlOneR}). Hence, we should only consider
the Ramond gauge symmetry, which is,
\begin{subequations}
\label{Rgaugeof1R}
\begin{equation}
\delta A=[\al,\chi]\,,\qquad \delta \al=\cQ \chi-\eta_0 \chi\,.
\end{equation}
The requirement that the gauge conditions are invariant under these
transformations are,
\begin{equation}
\label{1Rrestr}
[\al,\cQ \chi]=\eta_0 \cQ \chi=0\,.
\end{equation}
\end{subequations}
The last of these conditions can be solved resulting in,
\begin{equation}
\delta \al =\eta_0 \hat \chi\,.
\end{equation}
It seems reasonable that one can get this
expression by gauge fixing the gauge for gauge symmetry as,
\begin{equation}
\cQ \chi=0\,,
\end{equation}
which also completely resolves~(\ref{1Rrestr}).
Solving this condition leads exactly to~(\ref{NPRgaugeof1R})
with the identification,
\begin{equation}
\chi=\cQ \hat \La_{\frac{1}{2}}\,.
\end{equation}

\subsubsection{The non-polynomial RNS action}
\label{sec:NP_RNSaction}

For the purpose of deriving the equations of motion from the action, we have
to recast the action in terms of unconstrained string fields.
Using the gauge conditions~(\ref{pic0}),~(\ref{AofPhi})
and~(\ref{alGaugeCond}), the action~(\ref{RNSaction}) can be written as,
\begin{equation}
\label{RNSNPaction1}
S_{NP} = \oint A_\eta e^{-L} A_Q-\oint P \Big(\frac{1}{2}\al Q\al
   + e^{-\Phi}Q e^\Phi \al^2\Big).
\end{equation}
Here, we refrained from using $A_Q$ in the part that includes the $P$
mid-point insertion in order to avoid potential ambiguities.
The form of the action~(\ref{RNSNPaction1}) is still not satisfactory, since
it depends on the constrained string field $\al$. The problematic
constraint~(\ref{etaAl0}) can be resolved by reverting to the even $\Xi$
field~(\ref{alOfXi}). The resulting action,
\begin{equation}
\label{RNSNPaction}
S_{NP}=\oint A_\eta e^{-L} A_Q-\oint P \Big(\frac{1}{2}\eta_0 \Xi Q\eta_0 \Xi
   + e^{-\Phi}Q e^\Phi (\eta_0 \Xi)^2\Big),
\end{equation}
is unconstrained and can be used for deriving the equations of motion.
We can use the $\Ztwo$ symmetry, to be discussed in the following sections,
in order to define yet another fully RNS action,
\begin{equation}
\tilde S_{NP}=\oint A_Q e^{L} A_\eta
  -\oint \xi \Big(\frac{1}{2}Q \Xi \eta_0 Q \Xi
   + e^{\Phi}\eta_0 e^{-\Phi} (Q \Xi)^2\Big),
\end{equation}
where now $\Xi$ carries picture number $-\frac{1}{2}$.
This action should share the properties of~(\ref{RNSNPaction}). For
concreteness and in order to continue the discussion so far, we stick to the
action~(\ref{RNSNPaction}).

For deriving the equations of motion from~(\ref{RNSNPaction}), we have to
evaluate the variations with respect to $\Phi$ and $\Xi$.
The former variation leads to,
\begin{subequations}
\label{Rvar}
\begin{equation}
\delta_1 S_{NP}=\oint \Big[\big(\eta_0 A-\al^2\big)+
 P \cQ \big(\al^2\big)\Big]e^{-\Phi}\delta e^\Phi\,,
\end{equation}
while the latter leads to,
\begin{equation}
\delta_2 S_{NP}=\oint \Big[Y \cQ\al -P[\eta_0 A,\al]\Big]\delta \Xi\,.
\end{equation}
\end{subequations}
Here, we wrote the final expressions in terms of $A$ and $\al$, which
should be interpreted as functions of $\Phi$ and $\Xi$. The fact that we
used the unconstrained fields in the derivation of this expression implies
that the ``fundamental lemma of the calculus of variations'' holds and as a
result, the expressions inside the
square brackets vanish. These expressions include various distinct mid-point
insertions and would, thus, vanish, if and only if the coefficients of each
of those mid-point insertions separately vanish. The resulting equations of
motion are therefore,
\begin{subequations}
\begin{align}
\label{EOM_RNS_NP1}
\eta_0 A-\al^2 &=0\,,\\
\label{EOM_RNS_NP2}
\cQ \big(\al^2\big) &=0\,,\\
\label{EOM_RNS_NP3}
\cQ \al &=0\,,\\
\label{EOM_RNS_NP4}
[\eta_0 A,\al] &=0\,.
\end{align}
\end{subequations}
These equations are not independent, since~(\ref{EOM_RNS_NP2}) follows
from~(\ref{EOM_RNS_NP3}), while~(\ref{EOM_RNS_NP1})
implies~(\ref{EOM_RNS_NP4}).
Hence, the independent equations of motion are exactly those
of~(\ref{EOM1Rfield}), as predicted by Berkovits.

The transformations~(\ref{La0GaugeSym}) and
\begin{equation}
\delta \Xi=\eta_0 \hat \La_{3/2}\,,
\end{equation}
are easily recognized
as gauge symmetries of the action~(\ref{RNSNPaction}).
Of the other two expected transformations~(\ref{NPRgaugeof1R}), the one
associated with $\hat \La_1$ can also be seen to hold, e.g.,
using~(\ref{Rvar}).
The transformation induced by $\hat \La_{1/2}$, on the other hand, leaves
the action invariant only on-shell. It might be possible that some
generalization of it holds. However, it might also be the case that the
partial gauge fixing that we performed treats differently the on-shell and
off-shell cases. This seems plausible, since the various gauge symmetries
are intertwined in the original, democratic, theory.
At any rate, one can  try to further gauge fix the action~(\ref{RNSNPaction})
and use it as a starting point for deriving the RNS perturbation theory.
It would be important in this case to treat the mid-point insertion $P$ as
part of the definition of the R-sector measure, instead of trying to
``invert'' it. This might lead to some subtleties, similar to the ones
described in section 6 of~\cite{Kroyter:2009rn}. We leave the issues of
further gauge fixing and the construction of perturbation theory based on
this action to future work.

\section{Gauge fixing at picture number $-1$}
\label{sec:newSSFT}

Standard gauge fixing leads to the modified theory. This is dual to the
non-polynomial theory in the sense that the role of the gauge fixing and
the equation of motion is interchanged,
\begin{equation}
Q A+A^2=0\qquad \Longleftrightarrow \qquad \eta_0 A=0\,.
\end{equation}
There is also another duality, the $\Ztwo$ duality
of~\cite{Berkovits:1994vy,Berkovits:1996bf}, which we encounter below.

Recall first the consequences of fixing the theory to picture numbers zero
and $-\frac{1}{2}$.
Collecting the components of the equations of motion at different pictures
leads to,
\begin{subequations}
\label{zeroPicEOM}
\begin{align}
Q A + A^2 &= 0\,,\\
-\eta_0 A + \al^2 &=0\,,\\
Q\al +[A,\al] &= 0\,,\\
-\eta_0 \al &=0\,.
\end{align}
\end{subequations}
These equations are not quite those of the modified cubic theory. For one
thing, we separated terms at different picture numbers, while in the former
interpretation explicit mid-point insertions of picture changing operators
were used. However, there is also another difference: The ``gauge choice''
that fixes $A$ to the small space is inconsistent, since it would set
$\al^2=0$.
We shall not try here to resolve the Ramond sector gauge fixing associated
with the small space $pic(A)=0$ gauge choice, since we already gave an
alternative in the previous section. Instead we want to examine the NS sector
here and in the analogous $pic(A)=-1$ case.

Restricting to the NS sector, the equations of motion~(\ref{zeroPicEOM})
reduce to the familiar,
\begin{subequations}
\label{zeroPicNSEOM}
\begin{align}
Q A + A^2 &= 0\,,\\
\eta_0 A &=0\,.
\end{align}
\end{subequations}
Now, not only it is possible to fix the string field to the small
space, but it is forced upon us. The linearized gauge symmetry is,
\begin{subequations}
\label{zeroPicNSgauge}
\begin{align}
\delta A &= \cQ\La\,,\\
\eta_0 \La &=0\,.
\end{align}
\end{subequations}
Again, the familiar expression of the modified cubic theory.

Consider now the NS sector at $-1$ picture. The equations of motion are,
\begin{subequations}
\label{naturalPicNSEOM}
\begin{align}
\label{dualCond}
Q A &= 0\,,\\
-\eta_0 A + A^2 &=0\,.
\end{align}
\end{subequations}
These equations are identical to those of the previous case under a $\Ztwo$
symmetry that exchange $Q$ with $-\eta_0$. Alternatively, the $\Ztwo$ can
be interpreted as exchanging $Q$ with $\eta_0$, while sending $\Psi$ to
$-\Psi$. This is exactly the way this symmetry is realised in the
non-polynomial theory~\cite{Berkovits:1995ab}. However, that case is
different, since there the ghost and picture numbers of the string field
equal zero.

The equations of motion~(\ref{naturalPicNSEOM}) are very different from
those of Witten's theory.
This is a new theory, in which the string field is restricted to the dual
space. The cohomology of $\eta_0$ in the
dual space is the same as that of $Q$ in the usual small space.
If we take~(\ref{dualCond}) and $pic(A)=-1$ as the gauge fixing conditions
for the action~(\ref{action}) it reduces to,
\begin{equation}
\label{newAction}
S=\oint \cO_2 \Big(\frac{1}{2}A\eta_0 A+\frac{1}{3}A^3 \Big).
\end{equation}
Note, that we redefined $A \rightarrow -A$.
An alternative representation for the theory is,
\begin{equation}
\label{dualSpaceAction}
S=\dint X_2 \Big(\frac{1}{2}A\eta_0 A+\frac{1}{3}A^3 \Big),
\end{equation}
where the string field $A$ is implicitly assumed to live in the dual
space and where the new integration symbol means that the CFT
expectation value is to be evaluated in this space.
The double picture changing operator $X_2$ was obtained using
\begin{equation}
Q\cO_2 = X_2\,.
\end{equation}
In practice, it is more convenient to work with the
representation~(\ref{newAction}), since together with~(\ref{dualCond}) it
forms the simplest definition we have for~(\ref{dualSpaceAction}).

The linearized gauge symmetry takes the form,
\begin{subequations}
\label{naturalPicNSgauge}
\begin{align}
\delta A &= \eta_0\La+[A,\La]\,,\\
Q \La &=0\,.
\end{align}
\end{subequations}
Again, $\La$ is restricted to the dual space and the
gauge transformations take the expected form in this space.
The $\Ztwo$ symmetry between the new theory and the modified cubic theory is
naturally extended to cover the gauge symmetry.

As we mentioned in section~\ref{sec:GaugeFix}, an important property of the
modified cubic theory is the existence of the operator $c(0)\ket{0}$. Do we
have an analogue of this
operator in the new theory? Without it, it would be hard to believe that this
theory could be consistent even at the classical level with only the NS
string field. The existence of this operator is implied by the $\Ztwo$
symmetry. Indeed, a unique operator with picture number $-1$,
ghost number $1$ and conformal dimension $-1$ exists,
\begin{equation}
\label{hatC}
\hat c \equiv -\xi c c' e^{-2\phi}\,.
\end{equation}
The chosen normalization and sign are justified below by~(\ref{bcOPE}).
This operator belongs to the dual space in light of,
\begin{equation}
\hat c=Q(\xi c e^{-2\phi})\,.
\end{equation}
This is a reassuring evidence for the consistency
of the new theory. In section~\ref{sec:ErlerSol} below we show that
this operator plays an important role in the construction of a tachyon vacuum
solution.

\section{Analytical classical solutions of the new theory}
\label{sec:solutions}

We illustrate the classical consistency of the new theory by finding
analytical solutions thereof. We start with the case of marginal deformations
in~\ref{sec:Marginal} and follow with tachyon vacuum solutions
in~\ref{sec:ErlerSol}.

\subsection{Marginal deformation}
\label{sec:Marginal}

The first analytical solution describing marginal deformations were found
in~\cite{Schnabl:2007az,Kiermaier:2007ba}. Their RNS counterparts for the
non-polynomial theory were given
in~\cite{Erler:2007rh,Okawa:2007ri,Okawa:2007it}.
To get the analogue solutions in the modified cubic theory one can either
map these solutions using
(the mapping is described, e.g., in~\cite{Kling:2002vi,Fuchs:2008zx}),
\begin{equation}
A=e^{-\Phi}Q e^\Phi\,,
\end{equation}
or use some formal pure-gauge expression of the bosonic solution and
reinterpret them in the RNS theory.
A common limitation of the above constructions are that the marginal
deformations should have regular OPE's.

A construction of solutions for singular marginal deformations was given
in~\cite{Fuchs:2007yy,Fuchs:2007gw}. Another construction was given
in~\cite{Kiermaier:2007vu,Kiermaier:2007ki}. The former method relies on
the existence of a formal pure-gauge form for the solutions, while the
latter, which treats the singularities in a more systematic way, represents
the solutions in terms of the integrated vertex operators\footnote{A
novel representation, based on boundary changing operators was recently
presented in~\cite{Kiermaier:2010cf}.}. While it was not proven, we believe
that the two methods are equivalent~\cite{Fuchs:2008cc}. Here, we use the
former one and restrict ourselves to the photon marginal deformation for
simplicity.

In all the constructions, the solution was given as a sum,
\begin{equation}
\label{PsiN}
A=\sum_{n=1}^\infty \la^n A_n\,,
\end{equation}
where $\la$ is a parameter describing the strength of the marginal
deformation. The leading order term is known from CFT considerations,
\begin{equation}
\label{MarginalBC}
A_1=V(0)\ket{0},
\end{equation}
where $V$ is the unintegrated vertex operator associated with the marginal
deformation.

The building blocks of the construction~\cite{Fuchs:2007yy,Fuchs:2007gw},
were $V$ and its formal $Q$-primitive $W$, i.e.,
$W$ should obey,
\begin{equation}
\label{QWV}
Q W_{old} = V_{old}\,,
\end{equation}
but should not be considered as part of the Hilbert space. In our case, the
string field carries picture $-1$. Hence, it follows
from~(\ref{MarginalBC}) that $V$ should also carry this picture.
The unintegrated photon vertex operator in this picture is,
\begin{equation}
V=c\psi e^{-\phi}\,,\qquad \psi\equiv a_\mu \psi^\mu\,,
\end{equation}
and we assume that the vector $a$ was chosen such that,
\begin{equation}
\psi(z)\psi(0)\sim \frac{1}{z^2}\,.
\end{equation}
The case of a light-like deformation is simpler. One should merely follow
the discussion below, excluding the last step.

In order to derive $W$ we recall that the (integrand of the) integrated
vertex operator $U$ obeys,
\begin{equation}
Q U_{old} = \partial V_{old}\,,
\end{equation}
which together with~(\ref{QWV}) implies,
\begin{equation}
\partial W_{old}=U_{old}\,.
\end{equation}
The fact that we work in the dual space suggests that the equations that we
should solve are instead,
\begin{align}
\label{etaUdV}
\eta_0 U &=\partial V\,,\\
\partial W &=U\,.
\end{align}
In addition, $U$ should live in the dual space, i.e., it should obey,
\begin{equation}
\label{QU}
QU=0\,.
\end{equation}
The constraints~(\ref{etaUdV}) and~(\ref{QU}) together with the requirement
of a fixed picture number for $U$ fix it completely,
\begin{equation}
U=\partial(\xi c e^{-\phi} \psi)+i\partial X\,,\qquad X\equiv a_\mu X^\mu\,.
\end{equation}
From this expression we can read,
\begin{equation}
\label{W}
W=\xi c e^{-\phi} \psi+i X\,,
\end{equation}
and the formal nature of $W$ comes from its dependence on $x_0$, the zero
mode of $X(z)$, which is not defined if the space is compactified, i.e.,
if the deformation is indeed a non-trivial marginal deformation.

We write $U$ and $W$ in terms of $X(z)$, the holomorphic part of
$X(z,\bar z)$, since we would have to work with holomorphic
expressions\footnote{Strictly speaking, $X(z)$ is not holomorphic,
since it contains a logarithmic component. However, this component
drops out from all the computations relevant to the current discussion
and can be ignored. Everything works exactly the same as in the usual case,
described in~\cite{Fuchs:2007yy} and in 6.2 of~\cite{Fuchs:2008cc}.}.
Moreover, in this way we can avoid subtleties related to
boundary normal ordering.
In the general case, holomorphicity implies that the marginal deformation is
exactly marginal~\cite{Recknagel:1998ih}, as should be the case for the
photon.

The first term in the definition of $W$~(\ref{W}) has regular OPE with all
its powers. Hence, all the subtleties related to singularities of the OPE
come from the $X$ insertion. It follows that the singularity structure here
is identical to that of the photon solution of the modified theory.
Thus, we can immediately write down the solution,
\begin{subequations}
\label{LaMD}
\begin{align}
\label{MDsol}
A &=\eta_0 \La \frac{1}{1-\La}\,,\\
\La &=\sum_{n=1}^\infty \la^n \La_n\,,\\
\label{LaNmarg}
\La_n &=\frac{(-1)^{n-1}}{n!}(W^n\ket{0})\Omega^{n-1}\,.
\end{align}
\end{subequations}
The $W^n$ are implicitly normal ordered.

The proof that the solution~(\ref{MDsol}) is a genuine one, i.e., that
$A$ is $x_0$-independent, follows exactly as in the bosonic
case. We assume as the induction hypothesis that
$A_{<n}$ are $x_0$-independent. We then use~(\ref{PsiN}) and~(\ref{LaMD})
to write,
\begin{equation}
A_n=\eta_0 \La_n + \sum_{k=1}^{n-1} A_{n-k}\La_k\,.
\end{equation}
We see that,
\begin{equation}
\partial_{x_0} A_n=\eta_0 \partial_{x_0} \La_n
 + \sum_{k=1}^{n-1} A_{n-k}\partial_{x_0} \La_k\,,
\end{equation}
where we dropped out the terms that vanish according to the induction
hypothesis. One can see that this expression equals zero, provided that the
following holds,
\begin{equation}
\partial_{x_0} \La_n=-i\La_{n-1} \Omega\,.
\end{equation}
This identity follows immediately from the definition~(\ref{LaNmarg}).

In order to explicitly evaluate various coefficients, it might be
desirable to write the solution as a sum of fully normal ordered expressions.
To that end we write,
\begin{equation}
W^n=(i X)^n+n (i X)^{n-1} (\xi c e^{-\phi}\psi)
  +\frac{n(n-1)}{2}(i X)^{n-2} (\xi \xi' c c' e^{-2\phi})\,.
\end{equation}
The terms $\xi c e^{-\phi}\psi$ and $\xi \xi' c c' e^{-2\phi}$ are regular
and primary. The $X^n$ are not primary but their mutual OPE's can be deduced
from those of $e^{k X}$.
The absence of $x_0$ in the expression before normal ordering implies
its absence also in the fully normal ordered result.
From this fact it follows that the solution can be written in terms of integrals
of (powers of) $\partial X$.
We see that the solution now depends on $U$, just as
in~\cite{Kiermaier:2007vu}.

In some cases one might wish to impose the reality condition.
This can be achieved in a way that is very similar to
the bosonic case, i.e., we refer to the solution described so far as the
``left solution'' and define also the ``right solution''. These two solutions
are gauge equivalent and the explicit gauge transformation between the two
can be easily derived~\cite{Fuchs:2007yy}.
The real solution is then defined by going half-way along the gauge orbit
connecting the two solutions~\cite{Kiermaier:2007vu}.

\subsection{Tachyon vacuum solutions}
\label{sec:ErlerSol}

Schnabl's solution~\cite{Schnabl:2005gv} for the tachyon vacuum
of the bosonic theory\footnote{See
also~\cite{Okawa:2006vm,Fuchs:2006hw,Ellwood:2006ba,Bonora:2010hi}.},
was generalized by Erler to the case of the modified
theory~\cite{Erler:2007xt}. The
construction uses the formal pure gauge form of the solution and the
split string notations of~\cite{Erler:2006hw,Erler:2006ww}, which we
also employ. A $\Ztwo$ dual of this solution should exist in the dual theory.

Erler's solution is given in a formal gauge form using ``the same''
gauge string field $\La$ as in the bosonic case,
\begin{equation}
\La_E=B c \ket{0},
\end{equation}
with $B$ defined as,
\begin{equation}
B=-\int_{-i\infty}^{i \infty}b(z)dz\,.
\end{equation}
The solution itself is then given by,
\begin{equation}
A_E=Q\La_E\frac{1}{1-\La_E}\,.
\end{equation}

The most natural generalization for our case would be,
\begin{align}
\La=\hat B \hat c\ket{0}\,,\\
\label{Psi}
A=\eta_0 \La\frac{1}{1-\La}\,.
\end{align}
Again we want to have,
\begin{equation}
\hat B=-\int_{-i\infty}^{i \infty}\hat b(z)dz\,.
\end{equation}
What is needed, is the identification of the operators $\hat b$ and $\hat c$,
which for consistency should both live in the dual space.
We already identified $\hat c$ in~(\ref{hatC}).
The identification of $\hat b$ becomes simple when one examines the
$N=4$ generators of~\cite{Berkovits:1994vy} (eq. (5.1) there),
\begin{equation}
G^+ = J_B \,,\qquad \tilde G^+=\eta\,,\qquad G^- = b\,,
\end{equation}
where $J_B$ is (up to a total derivative) the BRST current.
From the above one can immediately deduce that our $\hat b$ should be the
$\tilde G^-$ generator of~\cite{Berkovits:1994vy}.
The simplest representation for this operator is using the identity,
\begin{equation}
\hat b=\tilde G^-=[G^+,J^{--}]=Q(b\xi)=\xi T_{tot}-bX+\xi''\,.
\end{equation}
In this representation it is clear that $b$ lives in the dual space,
since it is $Q$ exact in the large space.
It is also easy to see that another $N=4$ identity
holds,
\begin{equation}
\label{eta0HatB}
\eta_0 \hat b=T_{tot}\,.
\end{equation}
This is the analogy for our case of $Qb=T_{tot}$, which plays an important
role in Erler's construction. From~(\ref{eta0HatB}) we immediately see that,
\begin{equation}
\eta_0 \hat B=K\,,
\end{equation}
where $K$ equals that of the modified theory and thus requires no ``hat'',
\begin{equation}
K=-\int_{-i\infty}^{i \infty} T(z)dz\,.
\end{equation}
Another pair of identities that are easily verified are,
\begin{align}
\label{bcOPE}
\hat b(z) \hat c(0) &\sim \frac{1}{z}\,,\\
[\hat B,\hat c] &=1\,.
\end{align}

Substituting all the ingredients and using the relations they
obey,~(\ref{Psi}) reduces to,
\begin{equation}
A=F \hat c \frac{K \hat B}{1-\Omega}\hat c F
  +F(\hat c K \hat c-\eta_0 \hat c) \hat B F\,.
\end{equation}
Direct evaluation of the second term using the OPE leads to,
\begin{equation}
\label{HatcKc}
\hat c K \hat c-\eta_0 \hat c=c c' e^{-2\phi}
  +\frac{1}{12} \xi \xi' c c' c'' c''' e^{-4\phi}\,.
\end{equation}
One can see that this expression has regular OPE's with $\hat b$
and $\hat c$. Comparison with Erler's solution suggests that we should
think of this expression as $\hat \gamma^2$. We can make this statement more
precise. In the small space, $\gamma=\eta e^\phi$ is the unique universal,
i.e., matter-independent, operator with
conformal weight $-\frac{1}{2}$, ghost number one and picture number zero.
A permissible $\hat \gamma$ should also be universal, it should live in the
dual space and should have conformal weight $-\frac{1}{2}$, ghost number one
and picture number $-1$. These requirements fix $\hat \gamma$ up to
normalization,
\begin{equation}
\hat \gamma=c e^{-\phi}+\frac{1}{4} \xi \xi' c c' c'' e^{-3\phi}\,.
\end{equation}
Had we used just the inverse picture changing operator to find $\hat \gamma$,
we would have gotten only the first term. Indeed, for an on-shell tachyon
this would be enough and it is only on-shell vertices that can be manipulated
using picture changing operators. For general values of the momentum (in
flat space) our previous considerations imply that the $\Ztwo$ symmetry fixes
$\frac{1+2k^2}{4}$ as the coefficient of the second term. This expression
reduces to $\frac{1}{4}$ and $0$ for $k^2=0$ and $k^2=-\frac{1}{2}$
respectively.

Using the OPE again, one finds,
\begin{equation}
\hat \gamma^2 = c c' e^{-2\phi}
  +\frac{1}{12} \xi \xi' c c' c'' c''' e^{-4\phi}\,,
\end{equation}
in agreement with~(\ref{HatcKc}).
Then, evaluation of one further OPE gives the full algebraic structure,
\begin{equation}
\begin{aligned}
\eta_0 \hat B &=K\,, \qquad
 \eta_0 \hat c=\hat c K \hat c-\hat \gamma^2\,,\qquad
 \eta_0 \hat \gamma^2=\hat c K \hat \gamma^2-\hat \gamma^2 K \hat c\,,\\
\hat B^2 &=\hat c^2=[\hat \gamma^2,\hat c]=[\hat \gamma^2,\hat B]=0\,,\qquad
 [\hat B,\hat c]=1\,.
\end{aligned}
\end{equation}
We recognize that the algebraic structure is identical to the standard one.
Moreover, using it we see that the solution can now be also written in
a form, which is manifestly the $\Ztwo$ image of Erler's solution,
\begin{equation}
A=F \hat c \frac{K \hat B}{1-\Omega}\hat c F
  +F\hat \gamma^2 \hat B F\,.
\end{equation}

Instead of proving Sen's conjectures for this solution, we would
like to generalize it and prove the conjectures for a simpler solution.
One option for a generalization would be to study the one-parameter family of
solutions of~\cite{Aref'eva:2008ad,Fuchs:2008zx}.
However, if we want to get a simpler form of the solution, it would be
better to find the counterpart of~\cite{Gorbachev:2010zz} (see
also~\cite{Arroyo:2010fq}), which generalizes to the RNS case the simple
solutions of Erler and Schnabl~\cite{Erler:2009uj}.
The benefit of studying such solutions is that no phantom terms are needed
for the evaluation of the action. The new solution is,
\begin{equation}
\label{simpA}
A=\big(-\hat c+\eta_0(\hat B\hat c)\big)\frac{1}{1-K}\,.
\end{equation}
The equations of motion easily follow. The kinetic term is,
\begin{equation}
\eta_0 A =-\eta_0 \hat c\frac{1}{1-K}\,.
\end{equation}
Writing the interaction term as,
\begin{equation}
A^2 = \Big[\big(-\hat c+\eta_0(\hat B \hat c)\big)\frac{1}{1-K}
   \big(-\hat c+\eta_0(\hat B \hat c)\big)\Big]\frac{1}{1-K}\,,
\end{equation}
we see that the expression inside the square brackets equals,
\begin{equation}
\Big[\cdots\Big]=\big(\hat c+\eta_0(\hat c \hat B)\big)
    \Big(\hat c+\frac{\hat B}{1-K}\eta_0\hat c\Big)
 =\hat c\hat B \eta_0\hat c+\eta_0(\hat c \hat B)\hat c
 =\eta_0(\hat c\hat B\hat c)=\eta_0\hat c\,,
\end{equation}
as it should.

We now turn to proving Sen's conjectures, namely, we want to prove the
triviality of the cohomology around the solution and show that the action of
the solution equals minus the volume times the D-brane tension.

\subsubsection{Trivial cohomology}

The, by now standard~\cite{Ellwood:2001ig,Ellwood:2006ba,Erler:2007xt},
method for proving the triviality of the cohomology is to establish the
existence of a contracting homotopy operator, i.e., a string field $\cA$
obeying $\cQ\cA=1$, where $1$ is the identity string field.
Of course, in our case we have to show,
\begin{equation}
\eta_0 \cA+[A_0,\cA]=1\,.
\end{equation}
Again, $\cA$ can immediately be guessed using the $\Ztwo$ symmetry,
\begin{equation}
\cA=-\hat B \frac{1}{1-K}\,.
\end{equation}
Indeed,
\begin{equation}
\eta_0 \cA+[A_0,\cA]=\Big(-K+\hat c\hat B\frac{1}{1-K}
  +\frac{1}{1-K}\hat B\hat c-\hat c \frac{\hat B K}{1-K}
  -\frac{\hat B K}{1-K}\hat c\Big)\frac{1}{1-K}=1\,.
\end{equation}

\subsubsection{The action of the solution}

In order to evaluate the action, we could presumably use once again the
$\Ztwo$ symmetry, defining explicitly the expectation values in the dual
space for the $\hat b \hat c$ sector, for the $PQ$
($\hat\xi \hat\eta$) sector, etc.
To that end, one would have to define a few more dual operators and show
that the expectation values factor properly. Then, one would get to
expressions that precisely match those that were already evaluated in the
literature.
Instead of going in this direction, we would like to evaluate the expectation
values directly. The expressions that one sees in this way differ from those
found before. The end result is, however, the same: Sen's conjecture holds
for our solution.

The energy per unit volume, i.e., ignoring the $\delta(0)$ factor that comes
from the zero modes of the $X$ sector, should equal $-\frac{1}{2\pi^2}$.
Since our solution does not depend on the matter sector, we can simply
define the integration as the expectation value in the ghost sectors. Then,
we have to prove that,
\begin{equation}
\label{SenFirst}
E=-S=-\frac{1}{2\pi^2}\,.
\end{equation}

We can use the equations of motion to write,
\begin{equation}
\label{actionSol}
S=\frac{1}{6}\oint \cO_2 A \eta_0 A\,.
\end{equation}
The picture number two mid-point insertion $\cO_2$ could be any regularized
and primary version of $\xi X$. We choose,
\begin{equation}
\label{O2xiX}
\cO_2=\xi(i) X(-i)\,,
\end{equation}
where the $\pm i$ are upper half place coordinates.
It was shown in~\cite{Kroyter:2009rn} that the specific choice of $\cO$
does not change the value of the action for solutions. Moreover, for the
specific choice~(\ref{O2xiX}), one can exchange the location
of the two insertions even without using the fact that $A$ is a solution,
\begin{equation}
\oint \xi(i) X(-i) A \eta_0 A=
\oint \xi(i) Q \xi (-i) A \eta_0 A=
\oint X(i) \xi(-i) A \eta_0 A\,.
\end{equation}
Here, we used the definition of $X$, integrated by parts
and used the fact that $A$ lives in the dual space.

Substituting~(\ref{simpA}) in~(\ref{actionSol}) we get,
\begin{equation}
S=\frac{1}{6}\oint \cO_2
\big(\hat c-\eta_0(\hat B\hat c)\big)\frac{1}{1-K}\eta_0 \hat c\frac{1}{1-K}
   =\frac{1}{6}\oint \cO_2
    \hat c\frac{1}{1-K}\eta_0 \hat c\frac{1}{1-K}\,,
\end{equation}
where we dropped the second term, which is a total $\eta_0$ derivative in the
dual space. One could also understand why does this term vanish in
the large space by integrating the $\eta_0$ by parts, thus killing the
$\xi$ insertion, and then writing $X=Q\xi$ and integrating $Q$ by parts.
Next, we substitute,
\begin{equation}
\frac{1}{1-K}=\int_0^\infty e^{-t(1-K)}=\int_0^\infty e^{-t}W_t\,,
\end{equation}
where $W_t$ are wedge states.

We see that in contrast to many similar expressions, previously evaluated in
the literature, in the case at hand we are left with no line integrals.
Hence, we transform all expressions to the upper half-plane (in accord with
our use of $\pm i$ for the mid-point insertions), instead of evaluating the
expression in the cylinder coordinates with their funny looking correlators.
Both, $\hat c$ and $\eta_0 \hat c$ are primaries of weight $-1$.
In the cylinder coordinates these insertions are located at $0$ and $t$.
The conformal transformation to the upper half plane takes the form,
\begin{equation}
\zeta(z)=\tan\Big(\frac{\pi z}{t+s}\Big)\,,
\end{equation}
where $\zeta$ is the upper half plane coordinate and $z$ is the cylinder
coordinate, where the lines $\Re(z)=0$ and $\Re(z)=t+s$ are identified.
Applying the conformal transformation leads to,
\begin{equation}
\begin{aligned}
S=&\int_0^\infty dt\,ds\frac{e^{-t-s}(t+s)^2}{6\pi^2(1+\zeta(t)^2)}
  \cdot\\ \cdot &
 \Big<\xi(i)\big(2e^{2\phi}b\eta'+e^{2\phi}b'\eta+(e^{2\phi})'b\eta\big)(-i)
 \big(\xi c c' e^{-2\phi}\big)(0) \big(c c' e^{-2\phi}\big)(\zeta(t))\Big>
\end{aligned}
\end{equation}
We can write the expectation value as,
\begin{align}
\Big<\cdots\Big>=& \\ \partial_\zeta\circ & \nonumber \Big(
  \big<\xi(i)\eta(\zeta)\xi(0)\big>_{\xi \eta}
  \big<b(\zeta)(cc')(0)(cc')(\zeta(t))\big>_{bc}
  \big<e^{2\phi}(\zeta)e^{-2\phi}(0)e^{-2\phi}(\zeta(t))\big>_{\phi}  
\Big)\Big|_{\zeta=-i}\,.
\end{align}
Here, we define the derivative to act in the correct way for reproducing the
$X$ insertion, i.e., $\partial_\zeta\circ$ represents the sum of the
derivatives acting on the three correlators, with a factor of two multiplying
the one acting on the first correlator. Substitution of standard expressions
reveals\footnote{We follow the conventions $\vev{cc'c''}=2$, in accord with
our choice of a minus sign for the action~(\ref{action}).},
\begin{equation}
\Big<\cdots\Big>_{\xi \eta}=\frac{i}{\zeta(i-\zeta)}\,,\qquad
\Big<\cdots\Big>_{b c}=-
  \frac{\zeta(t)^4}{\zeta^2 \big(\zeta(t)-\zeta\big)^2}\,,\qquad
\Big<\cdots\Big>_{\phi}=
  \frac{\zeta^4 \big(\zeta(t)-\zeta\big)^4}{\zeta(t)^4}\,.
\end{equation}
Applying the derivative and putting it all together leads to a very simple
expression,
\begin{equation}
S=\int_0^\infty \frac{e^{-t-s}(t+s)^2}{12\pi^2}\,dt\,ds=
  \frac{1}{2\pi^2}\,,
\end{equation}
in agreement with~(\ref{SenFirst}).

\section{Conclusions}
\label{sec:conc}

In this paper we demonstrated the consistency of the democratic theory by
showing that it can be reduced to the reliable non-polynomial theory by a
specific partial gauge fixing.
Moreover, we managed to extend this partial gauge fixing and obtained a
string field theory action, at fixed picture numbers for the open RNS string.
We further explained, that in contrast with some naive expectations, the
democratic theory cannot be gauge fixed to produce
Witten's theory. A gauge fixing to the $-1$ picture of the NS string field is
possible. However, it leads to a new theory, which is the $\Ztwo$ counterpart
of the modified theory. The $\Ztwo$ symmetry was also used for generating
another variant of the non-polynomial RNS string field theory action.

One could criticize our construction on several grounds. An obvious criticism
is that we did not prove that our gauge choice~(\ref{AofPhi}) is globally
permissible. On the other hand, we do not know also whether Siegel's gauge
or Schnabl's gauge are globally permissible. We do know that at the
linearized level (with respect to the string field) our choice is adequate.
Our expectation is that if there are issues related to its global validity, they
would probably imply that some legitimate configurations become singular when
represented in the non-polynomial theory and not that there is a problem with
the democratic theory itself.

The problem with showing that any of our gauge choices is globally
permissible is also related to the problem of defining the space of string
fields. This is another potential source of criticism on our construction.
While there is a lack of understanding regarding the definition of these
spaces in all existing variants of string field theory, there are two
specific points that are particular to our construction.
The first issue is related to
the use of mid-point insertions in the action, which we resolved by demanding
that the space of string fields does not include string fields that can be
interpreted as having mid-point insertions. This resolution is most
probably correct. The reason is that if it does not hold, then singularities
would emerge from star multiplying string fields in general, regardless of
the existence of mid-point insertions or their regularity. Thus, if this
constraint cannot be enforced, then all string field theories that rely
on Witten's star product ought to be inconsistent, i.e., not only our theory,
but also those with ``regular'' mid-point insertions in the
action~\cite{Berkovits:2005bt,Berkovits:2009gi,Kroyter:2009zj}
and those with actions that lack mid-point insertions
altogether~\cite{Witten:1986cc,Berkovits:1995ab}.
An encouraging observation is that the algebra of finite star products of
states with no mid-point insertions closes on states with no mid-point
insertions\footnote{An alternative to the constructions based on the star
product exists, namely the one that includes stubs, similarly to the case of
closed string field theory~\cite{Zwiebach:1992ie}. Constructions of
stub-based open and open-closed string field theories appeared
in~\cite{Zwiebach:1992bw,Zwiebach:1997fe}. These theories tend to be
non-polynomial.}.

The other delicate point regarding the space of string fields is our
requirement that the string field components decay fast enough as a
function of the picture number. Again, we do not know how to quantify this
condition, except for vertex operators. For those we know that we are
dealing with an infinite number of copies of the same object. We therefore
have to require that the sum of coefficients (as calculated from any base
representation using multi-picture changing operators and while ignoring
exact terms) of any vertex operator is absolutely convergent. This condition
invalidates the use of the
contracting homotopy operators of~\cite{Berkovits:2001us} and
leads to a correct cohomology problem~\cite{Kroyter:2009rn}. We believe that some
sort of a generalization of this condition to off-shell states should exist.
What could have resolved this issue is a positive definite norm that would have
allowed us to compare the ``size'' of different vertex operators and to include
also non-closed states. Unfortunately, a canonical norm of this sort does not exist.
The lack of the norm and the inability to compare different vertex operators
is exactly the usual problem with defining the space of
string fields. Thus, as usual, we do not have a definition for the desired space,
but we know some properties thereof and it seems that a proper definition
should exist. This is in contrast to, e.g., the situation
with the pure-spinor string field theory, where it seems that there is no
hope of obtaining a sensible space of string fields using the current
formulation of the theory~\cite{Aisaka:2008vw,Bedoya:2009np}\footnote{It should be
noted, however, that this theory makes at least some sense at the classical
level, since it supports analytical solutions for marginal
deformations~\cite{kroyter:2010xx}.}.

Another ground for criticizing the democratic theory could be the presence
of operators of arbitrarily negative conformal weight. We believe that this
should not be considered a problem of principle, since these operators
correspond to auxiliary fields. Nonetheless, this state of affairs can pose
a difficulty to some numerical, e.g., lattice~\cite{Bursa:2010sg} studies
of the theory.
One might try to overcome this by modifying the democratic theory into
a ``semi-democratic'' one along the lines of ``the big
picture''~\cite{Berkovits:1991gj}. It might be interesting to study
this possibility.

There is still much more to be done regarding the formulation of the
democratic theory. Other than the issues mentioned already the most salient
point is the understanding of supersymmetry. Despite some ideas presented
in~\cite{Kroyter:2009rn}, we don't know if and how does supersymmetry act
on off-shell states. Note, that this is not the ``usual'' problem with
supersymmetry that does not close off-shell. Here, we don't even know how
to define it off-shell. Hence, we cannot even claim that it is a symmetry.
The problem with the construction of supersymmetry
is that singular OPEs can occur between the mid-point insertion $\cO$ and
the supersymmetry current, regardless of its picture. It might happen that a
specific combination of pictures for the current resolves this problem.
Another possibility is to use the freedom of adding exact terms to $\cO$.
While these terms are of no
importance on-shell, they might lead to different results off-shell.
It would be interesting to find out whether the construction of supersymmetry
constraints these terms. While such a restriction on the form
of the exact terms would come from supersymmetry, the derived form of
$\cO$ would be universal.

There are several other challenges and possible directions for future
research. Devising other gauge fixings, in particular fixings that do not
include a projection to given picture numbers (other than the trivial cases
mentioned in~(\ref{HLa})) would be very interesting.
Deriving the surface states associated with a classical
solutions along the lines of~\cite{Kiermaier:2008qu} would also be useful.
Generalization of the construction to some of the other theories studied
in~\cite{Berkovits:1994vy} by replacing $Q$ and $\eta_0$ by the more general
$G^\pm$ might lead to string field theories around new backgrounds.
In the cases in which these operators possess non-trivial cohomologies
there would be no contracting homotopy
operators and hence no picture changing operators. There would be no
gauge symmetry associated with picture changing, still the action might be
correct, provided that the $\cO$ insertion can be found. A related issue is
the understanding of the correspondence, if any, between the democratic
theory and pure-spinor string field theory.

Another possible research direction relates to the recent observation
that the modified theory and the non-polynomial theory support
different classes of classical solutions~\cite{Erler:2010pr}, at least
if one does not impose the reality condition.
This observation appears to contradict our claim that both these theories
can be obtained from the NS sector of the democratic theory by a partial
gauge fixing.
It seems that this contradiction can be resolved in one of a few ways.
It might be the case that one of the gauge fixings that we employed breaks
down at some finite value of the string field. 
Another possibility is that not all of the assumptions of~\cite{Erler:2010pr}
hold, e.g., it might be the case that the $\cL^-$ expansion is not a
legitimate one or that the $c,B,K,G,\gamma$ subalgebra considered there is
essentially different from the complete string field algebra.
At any rate, we believe that the democratic theory before gauge fixing is
the more reliable one. The resolution of this puzzle
might shed new light on issues such as gauge fixing in string field theory
and the construction of string field spaces.

As a last idea regarding future directions we would like to suggest the
construction of heterotic and closed RNS string field theories along the
lines of the democratic theory.
Explicit insertions of picture changing operators might be useful for
resolving the difficulties with the Ramond sectors of these theories.
This idea gets complicated by the fact that closed strings have no
``mid-points''. Nonetheless, one can look for other special points for
the insertion of operators.
In fact, such a construction was already carried out successfully
by Saroja and Sen~\cite{Saroja:1992vw}. However, it was limited to the NS
sector. It might well be the case that adding some of the element of the
democratic theory to their construction would allow for the inclusion of
the Ramond sectors. We are currently studying this issue.

\section*{Acknowledgments}

I would like to thank Nathan Berkovits, Stefan Fredenhagen, Udi Fuchs,
Michael Kiermaier, Martin Schnabl, Cobi Sonnenschein, Leonardo Rastelli,
Adam Schwimmer, Warren Siegel and Barton Zwiebach for discussions.

This work was supported by the U.S. Department of Energy
(D.O.E.) under cooperative research agreement DE-FG0205ER41360.
My research is supported by an Outgoing International Marie Curie
Fellowship of the European Community. The views presented in this work are
those of the author and do not necessarily reflect those of the European
Community.

\bibliography{bib}

\providecommand{\href}[2]{#2}\begingroup\raggedright\begin{thebibliography}{10}

\bibitem{Sen:1999mh}
A.~Sen, {\it Descent relations among bosonic {D}-branes},  {\em Int. J. Mod.
  Phys.} {\bf A14} (1999) 4061--4078,
  [\href{http://xxx.lanl.gov/abs/hep-th/9902105}{{\tt hep-th/9902105}}].

\bibitem{Sen:1999xm}
A.~Sen, {\it Universality of the tachyon potential},  {\em JHEP} {\bf 12}
  (1999) 027, [\href{http://xxx.lanl.gov/abs/hep-th/9911116}{{\tt
  hep-th/9911116}}].

\bibitem{Fuchs:2008cc}
E.~Fuchs and M.~Kroyter, {\it Analytical solutions of open string field
  theory},  \href{http://xxx.lanl.gov/abs/0807.4722}{{\tt 0807.4722}}.

\bibitem{Schnabl:2010tb}
M.~Schnabl, {\it Algebraic solutions in open string field theory - a lightning
  review},  \href{http://xxx.lanl.gov/abs/1004.4858}{{\tt 1004.4858}}.

\bibitem{Kroyter:2009rn}
M.~Kroyter, {\it Superstring field theory in the democratic picture},
  \href{http://xxx.lanl.gov/abs/0911.2962}{{\tt 0911.2962}}.

\bibitem{Friedan:1985ge}
D.~Friedan, E.~J. Martinec, and S.~H. Shenker, {\it Conformal invariance,
  supersymmetry and string theory},  {\em Nucl. Phys.} {\bf B271} (1986) 93.

\bibitem{Horowitz:1988ip}
G.~T. Horowitz, R.~C. Myers, and S.~P. Martin, {\it {BRST} cohomology of the
  superstring at arbitrary ghost number},  {\em Phys. Lett.} {\bf B218} (1989)
  309.

\bibitem{Lian:1989cy}
B.~H. Lian and G.~J. Zuckerman, {\it {BRST} cohomology of the supervirasoro
  algebras},  {\em Commun. Math. Phys.} {\bf 125} (1989) 301.

\bibitem{Berkovits:1994vy}
N.~Berkovits and C.~Vafa, {\it {N=4 topological strings}},  {\em Nucl. Phys.}
  {\bf B433} (1995) 123--180,
  [\href{http://xxx.lanl.gov/abs/hep-th/9407190}{{\tt hep-th/9407190}}].

\bibitem{Berkovits:1996bf}
N.~Berkovits, {\it A new description of the superstring},
  \href{http://xxx.lanl.gov/abs/hep-th/9604123}{{\tt hep-th/9604123}}.

\bibitem{Wendt:1987zh}
C.~Wendt, {\it Scattering amplitudes and contact interactions in {W}itten's
  superstring field theory},  {\em Nucl. Phys.} {\bf B314} (1989) 209.

\bibitem{Berkovits:1995ab}
N.~Berkovits, {\it Super-{P}oincar\'e invariant superstring field theory},
  {\em Nucl. Phys.} {\bf B450} (1995) 90--102,
  [\href{http://xxx.lanl.gov/abs/hep-th/9503099}{{\tt hep-th/9503099}}].

\bibitem{Arefeva:1989cm}
I.~Y. Arefeva, P.~B. Medvedev, and A.~P. Zubarev, {\it Background formalism for
  superstring field theory},  {\em Phys. Lett.} {\bf B240} (1990) 356--362.

\bibitem{Arefeva:1989cp}
I.~Y. Arefeva, P.~B. Medvedev, and A.~P. Zubarev, {\it New representation for
  string field solves the consistency problem for open superstring field
  theory},  {\em Nucl. Phys.} {\bf B341} (1990) 464--498.

\bibitem{Preitschopf:1989fc}
C.~R. Preitschopf, C.~B. Thorn, and S.~A. Yost, {\it Superstring field theory},
   {\em Nucl. Phys.} {\bf B337} (1990) 363--433.

\bibitem{Kroyter:2009zi}
M.~Kroyter, {\it Superstring field theory equivalence: {Ramond} sector},  {\em
  JHEP} {\bf 10} (2009) 044, [\href{http://xxx.lanl.gov/abs/0905.1168}{{\tt
  0905.1168}}].

\bibitem{Berkovits:2001im}
N.~Berkovits, {\it The {R}amond sector of open superstring field theory},  {\em
  JHEP} {\bf 11} (2001) 047,
  [\href{http://xxx.lanl.gov/abs/hep-th/0109100}{{\tt hep-th/0109100}}].

\bibitem{Michishita:2004by}
Y.~Michishita, {\it A covariant action with a constraint and {F}eynman rules
  for fermions in open superstring field theory},  {\em JHEP} {\bf 01} (2005)
  012, [\href{http://xxx.lanl.gov/abs/hep-th/0412215}{{\tt hep-th/0412215}}].

\bibitem{Berkovits:2001us}
N.~Berkovits, {\it Relating the {RNS} and pure spinor formalisms for the
  superstring},  {\em JHEP} {\bf 08} (2001) 026,
  [\href{http://xxx.lanl.gov/abs/hep-th/0104247}{{\tt hep-th/0104247}}].

\bibitem{Kroyter:2009zj}
M.~Kroyter, {\it On string fields and superstring field theories},  {\em JHEP}
  {\bf 08} (2009) 044, [\href{http://xxx.lanl.gov/abs/0905.1170}{{\tt
  0905.1170}}].

\bibitem{Witten:1986cc}
E.~Witten, {\it Noncommutative geometry and string field theory},  {\em Nucl.
  Phys.} {\bf B268} (1986) 253.

\bibitem{Berkovits:2000hf}
N.~Berkovits, A.~Sen, and B.~Zwiebach, {\it Tachyon condensation in superstring
  field theory},  {\em Nucl. Phys.} {\bf B587} (2000) 147--178,
  [\href{http://xxx.lanl.gov/abs/hep-th/0002211}{{\tt hep-th/0002211}}].

\bibitem{Kroyter:2009bg}
M.~Kroyter, {\it Comments on superstring field theory and its vacuum solution},
   {\em JHEP} {\bf 08} (2009) 048,
  [\href{http://xxx.lanl.gov/abs/0905.3501}{{\tt 0905.3501}}].

\bibitem{Henneaux:1992ig}
M.~Henneaux and C.~Teitelboim, {\it Quantization of gauge systems}, .
  Princeton, USA: Univ. Pr. (1992) 520 p.

\bibitem{DeSmet:2000je}
P.-J. De~Smet and J.~Raeymaekers, {\it The tachyon potential in {W}itten's
  superstring field theory},  {\em JHEP} {\bf 08} (2000) 020,
  [\href{http://xxx.lanl.gov/abs/hep-th/0004112}{{\tt hep-th/0004112}}].

\bibitem{Arefeva:2002mb}
I.~Y. Arefeva, D.~M. Belov, and A.~A. Giryavets, {\it Construction of the
  vacuum string field theory on a non-{BPS} brane},  {\em JHEP} {\bf 09} (2002)
  050, [\href{http://xxx.lanl.gov/abs/hep-th/0201197}{{\tt hep-th/0201197}}].

\bibitem{Aref'eva:2000mb}
I.~Y. Aref'eva, A.~S. Koshelev, D.~M. Belov, and P.~B. Medvedev, {\it Tachyon
  condensation in cubic superstring field theory},  {\em Nucl. Phys.} {\bf
  B638} (2002) 3--20, [\href{http://xxx.lanl.gov/abs/hep-th/0011117}{{\tt
  hep-th/0011117}}].

\bibitem{Gates:1983nr}
S.~J. Gates, M.~T. Grisaru, M.~Rocek, and W.~Siegel, {\it Superspace, or one
  thousand and one lessons in supersymmetry},  {\em Front. Phys.} {\bf 58}
  (1983) 1--548, [\href{http://xxx.lanl.gov/abs/hep-th/0108200}{{\tt
  hep-th/0108200}}].

\bibitem{Horowitz:1986dt}
G.~T. Horowitz, J.~Lykken, R.~Rohm, and A.~Strominger, {\it A purely cubic
  action for string field theory},  {\em Phys. Rev. Lett.} {\bf 57} (1986)
  283--286.

\bibitem{Berkovits:2004xh}
N.~Berkovits, Y.~Okawa, and B.~Zwiebach, {\it {WZW}-like action for heterotic
  string field theory},  {\em JHEP} {\bf 11} (2004) 038,
  [\href{http://xxx.lanl.gov/abs/hep-th/0409018}{{\tt hep-th/0409018}}].

\bibitem{Witten:1983ar}
E.~Witten, {\it Nonabelian bosonization in two-dimensions},  {\em
  Commun.Math.Phys.} {\bf 92} (1984) 455--472.

\bibitem{Henneaux:1989jq}
M.~Henneaux, {\it Lectures on the {Antifield-BRST} formalism for gauge
  theories},  {\em Nucl. Phys. Proc. Suppl.} {\bf 18A} (1990) 47--106.

\bibitem{Gomis:1994he}
J.~Gomis, J.~Paris, and S.~Samuel, {\it Antibracket, antifields and gauge
  theory quantization},  {\em Phys. Rept.} {\bf 259} (1995) 1--145,
  [\href{http://xxx.lanl.gov/abs/hep-th/9412228}{{\tt hep-th/9412228}}].

\bibitem{Schnabl:2007az}
M.~Schnabl, {\it Comments on marginal deformations in open string field
  theory},  {\em Phys. Lett.} {\bf B654} (2007) 194--199,
  [\href{http://xxx.lanl.gov/abs/hep-th/0701248}{{\tt hep-th/0701248}}].

\bibitem{Kiermaier:2007ba}
M.~Kiermaier, Y.~Okawa, L.~Rastelli, and B.~Zwiebach, {\it Analytic solutions
  for marginal deformations in open string field theory},  {\em JHEP} {\bf 01}
  (2008) 028, [\href{http://xxx.lanl.gov/abs/hep-th/0701249}{{\tt
  hep-th/0701249}}].

\bibitem{Erler:2007rh}
T.~Erler, {\it Marginal solutions for the superstring},  {\em JHEP} {\bf 07}
  (2007) 050, [\href{http://xxx.lanl.gov/abs/arXiv:0704.0930 [hep-th]}{{\tt
  arXiv:0704.0930 [hep-th]}}].

\bibitem{Okawa:2007ri}
Y.~Okawa, {\it Analytic solutions for marginal deformations in open superstring
  field theory},  {\em JHEP} {\bf 09} (2007) 084,
  [\href{http://xxx.lanl.gov/abs/arXiv:0704.0936 [hep-th]}{{\tt arXiv:0704.0936
  [hep-th]}}].

\bibitem{Okawa:2007it}
Y.~Okawa, {\it Real analytic solutions for marginal deformations in open
  superstring field theory},  {\em JHEP} {\bf 09} (2007) 082,
  [\href{http://xxx.lanl.gov/abs/arXiv:0704.3612 [hep-th]}{{\tt arXiv:0704.3612
  [hep-th]}}].

\bibitem{Kling:2002vi}
A.~Kling, O.~Lechtenfeld, A.~D. Popov, and S.~Uhlmann, {\it Solving string
  field equations: {New} uses for old tools},  {\em Fortsch. Phys.} {\bf 51}
  (2003) 775--780, [\href{http://xxx.lanl.gov/abs/hep-th/0212335}{{\tt
  hep-th/0212335}}].

\bibitem{Fuchs:2008zx}
E.~Fuchs and M.~Kroyter, {\it On the classical equivalence of superstring field
  theories},  {\em JHEP} {\bf 10} (2008) 054,
  [\href{http://xxx.lanl.gov/abs/0805.4386}{{\tt 0805.4386}}].

\bibitem{Fuchs:2007yy}
E.~Fuchs, M.~Kroyter, and R.~Potting, {\it Marginal deformations in string
  field theory},  {\em JHEP} {\bf 09} (2007) 101,
  [\href{http://xxx.lanl.gov/abs/arXiv:0704.2222 [hep-th]}{{\tt arXiv:0704.2222
  [hep-th]}}].

\bibitem{Fuchs:2007gw}
E.~Fuchs and M.~Kroyter, {\it Marginal deformation for the photon in
  superstring field theory},  {\em JHEP} {\bf 11} (2007) 005,
  [\href{http://xxx.lanl.gov/abs/arXiv:0706.0717 [hep-th]}{{\tt arXiv:0706.0717
  [hep-th]}}].

\bibitem{Kiermaier:2007vu}
M.~Kiermaier and Y.~Okawa, {\it Exact marginality in open string field theory:
  a general framework},  {\em JHEP} {\bf 11} (2009) 041,
  [\href{http://xxx.lanl.gov/abs/0707.4472}{{\tt 0707.4472}}].

\bibitem{Kiermaier:2007ki}
M.~Kiermaier and Y.~Okawa, {\it General marginal deformations in open
  superstring field theory},  {\em JHEP} {\bf 11} (2009) 042,
  [\href{http://xxx.lanl.gov/abs/0708.3394}{{\tt 0708.3394}}].

\bibitem{Kiermaier:2010cf}
M.~Kiermaier, Y.~Okawa, and P.~Soler, {\it Solutions from boundary condition
  changing operators in open string field theory},
  \href{http://xxx.lanl.gov/abs/1009.6185}{{\tt 1009.6185}}.

\bibitem{Recknagel:1998ih}
A.~Recknagel and V.~Schomerus, {\it Boundary deformation theory and moduli
  spaces of {D}-branes},  {\em Nucl. Phys.} {\bf B545} (1999) 233--282,
  [\href{http://xxx.lanl.gov/abs/hep-th/9811237}{{\tt hep-th/9811237}}].

\bibitem{Schnabl:2005gv}
M.~Schnabl, {\it Analytic solution for tachyon condensation in open string
  field theory},  {\em Adv. Theor. Math. Phys.} {\bf 10} (2006) 433--501,
  [\href{http://xxx.lanl.gov/abs/hep-th/0511286}{{\tt hep-th/0511286}}].

\bibitem{Okawa:2006vm}
Y.~Okawa, {\it Comments on {S}chnabl's analytic solution for tachyon
  condensation in {W}itten's open string field theory},  {\em JHEP} {\bf 04}
  (2006) 055, [\href{http://xxx.lanl.gov/abs/hep-th/0603159}{{\tt
  hep-th/0603159}}].

\bibitem{Fuchs:2006hw}
E.~Fuchs and M.~Kroyter, {\it On the validity of the solution of string field
  theory},  {\em JHEP} {\bf 05} (2006) 006,
  [\href{http://xxx.lanl.gov/abs/hep-th/0603195}{{\tt hep-th/0603195}}].

\bibitem{Ellwood:2006ba}
I.~Ellwood and M.~Schnabl, {\it Proof of vanishing cohomology at the tachyon
  vacuum},  {\em JHEP} {\bf 02} (2007) 096,
  [\href{http://xxx.lanl.gov/abs/hep-th/0606142}{{\tt hep-th/0606142}}].

\bibitem{Bonora:2010hi}
L.~Bonora, C.~Maccaferri, and D.~D. Tolla, {\it Relevant deformations in open
  string field theory: {A} simple solution for lumps},
  \href{http://xxx.lanl.gov/abs/1009.4158}{{\tt 1009.4158}}.

\bibitem{Erler:2007xt}
T.~Erler, {\it Tachyon vacuum in cubic superstring field theory},  {\em JHEP}
  {\bf 01} (2008) 013, [\href{http://xxx.lanl.gov/abs/0707.4591}{{\tt
  0707.4591}}].

\bibitem{Erler:2006hw}
T.~Erler, {\it Split string formalism and the closed string vacuum},  {\em
  JHEP} {\bf 05} (2007) 083,
  [\href{http://xxx.lanl.gov/abs/hep-th/0611200}{{\tt hep-th/0611200}}].

\bibitem{Erler:2006ww}
T.~Erler, {\it Split string formalism and the closed string vacuum. {II}},
  {\em JHEP} {\bf 05} (2007) 084,
  [\href{http://xxx.lanl.gov/abs/hep-th/0612050}{{\tt hep-th/0612050}}].

\bibitem{Aref'eva:2008ad}
I.~Y. Aref'eva, R.~V. Gorbachev, and P.~B. Medvedev, {\it Tachyon solution in
  cubic {Neveu-Schwarz} string field theory},  {\em Theor. Math. Phys.} {\bf
  158} (2009) 320--332, [\href{http://xxx.lanl.gov/abs/0804.2017}{{\tt
  0804.2017}}].

\bibitem{Gorbachev:2010zz}
R.~V. Gorbachev, {\it New solution of the superstring equation of motion},
  {\em Theor. Math. Phys.} {\bf 162} (2010) 90--94.

\bibitem{Arroyo:2010fq}
E.~A. Arroyo, {\it Generating {Erler-Schnabl-type} solution for tachyon vacuum
  in cubic superstring field theory},  {\em J. Phys.} {\bf A43} (2010) 445403,
  [\href{http://xxx.lanl.gov/abs/1004.3030}{{\tt 1004.3030}}].

\bibitem{Erler:2009uj}
T.~Erler and M.~Schnabl, {\it A simple analytic solution for tachyon
  condensation},  {\em JHEP} {\bf 10} (2009) 066,
  [\href{http://xxx.lanl.gov/abs/0906.0979}{{\tt 0906.0979}}].

\bibitem{Ellwood:2001ig}
I.~Ellwood, B.~Feng, Y.-H. He, and N.~Moeller, {\it The identity string field
  and the tachyon vacuum},  {\em JHEP} {\bf 07} (2001) 016,
  [\href{http://xxx.lanl.gov/abs/hep-th/0105024}{{\tt hep-th/0105024}}].

\bibitem{Berkovits:2005bt}
N.~Berkovits, {\it Pure spinor formalism as an {N = 2} topological string},
  {\em JHEP} {\bf 10} (2005) 089,
  [\href{http://xxx.lanl.gov/abs/hep-th/0509120}{{\tt hep-th/0509120}}].

\bibitem{Berkovits:2009gi}
N.~Berkovits and W.~Siegel, {\it Regularizing cubic open {Neveu-Schwarz} string
  field theory},  {\em JHEP} {\bf 11} (2009) 021,
  [\href{http://xxx.lanl.gov/abs/0901.3386}{{\tt 0901.3386}}].

\bibitem{Zwiebach:1992ie}
B.~Zwiebach, {\it Closed string field theory: {Q}uantum action and the {B-V}
  master equation},  {\em Nucl. Phys.} {\bf B390} (1993) 33--152,
  [\href{http://xxx.lanl.gov/abs/hep-th/9206084}{{\tt hep-th/9206084}}].

\bibitem{Zwiebach:1992bw}
B.~Zwiebach, {\it Interpolating string field theories},  {\em Mod. Phys. Lett.}
  {\bf A7} (1992) 1079--1090,
  [\href{http://xxx.lanl.gov/abs/hep-th/9202015}{{\tt hep-th/9202015}}].

\bibitem{Zwiebach:1997fe}
B.~Zwiebach, {\it Oriented open-closed string theory revisited},  {\em Annals
  Phys.} {\bf 267} (1998) 193--248,
  [\href{http://xxx.lanl.gov/abs/hep-th/9705241}{{\tt hep-th/9705241}}].

\bibitem{Aisaka:2008vw}
Y.~Aisaka, E.~A. Arroyo, N.~Berkovits, and N.~Nekrasov, {\it Pure spinor
  partition function and the massive superstring spectrum},  {\em JHEP} {\bf
  08} (2008) 050, [\href{http://xxx.lanl.gov/abs/0806.0584}{{\tt 0806.0584}}].

\bibitem{Bedoya:2009np}
O.~A. Bedoya and N.~Berkovits, {\it {GGI} lectures on the pure spinor formalism
  of the superstring},  \href{http://xxx.lanl.gov/abs/0910.2254}{{\tt
  0910.2254}}.

\bibitem{kroyter:2010xx}
M.~Kroyter, {\it Analytical solutions of pure-spinor superstring field theory,
  to be published}, .

\bibitem{Bursa:2010sg}
F.~Bursa and M.~Kroyter, {\it Lattice string field theory},
  \href{http://xxx.lanl.gov/abs/1009.4414}{{\tt 1009.4414}}.

\bibitem{Berkovits:1991gj}
N.~Berkovits, M.~T. Hatsuda, and W.~Siegel, {\it The big picture},  {\em Nucl.
  Phys.} {\bf B371} (1992) 434--466,
  [\href{http://xxx.lanl.gov/abs/hep-th/9108021}{{\tt hep-th/9108021}}].

\bibitem{Kiermaier:2008qu}
M.~Kiermaier, Y.~Okawa, and B.~Zwiebach, {\it The boundary state from open
  string fields},  \href{http://xxx.lanl.gov/abs/0810.1737}{{\tt 0810.1737}}.

\bibitem{Erler:2010pr}
T.~Erler, {\it Exotic universal solutions in cubic superstring field theory},
  \href{http://xxx.lanl.gov/abs/1009.1865}{{\tt 1009.1865}}.

\bibitem{Saroja:1992vw}
R.~Saroja and A.~Sen, {\it Picture changing operators in closed fermionic
  string field theory},  {\em Phys. Lett.} {\bf B286} (1992) 256--264,
  [\href{http://xxx.lanl.gov/abs/hep-th/9202087}{{\tt hep-th/9202087}}].

\end{thebibliography}\endgroup

\end{document}